\documentclass[authordate1-4]{article}
\usepackage{natbib}

\usepackage{bm}
\usepackage{amsmath}

\usepackage{amsfonts}
\usepackage{graphicx}
\usepackage[outdir=./]{epstopdf}
\usepackage{afterpage}
\usepackage{url}
\usepackage{caption}

\usepackage[top=0.95in, bottom=1in, left=1.2in, right=1.2in]{geometry}
\usepackage{xcolor}
\newcommand{\bfS}{\textbf{S}}
\newcommand{\bfH}{\textbf{H}}
\newcommand{\bfR}{\textbf{R}}
\newcommand{\bfC}{\textbf{C}}
\newcommand{\bfD}{\textbf{D}}
\newcommand{\bfB}{\textbf{B}}

\newcommand{\bfx}{\textbf{x}}
\newcommand{\bfy}{\textbf{y}}

\newcommand{\Rp}{\bfR_{RR}}

\newcommand{\Rctrl}{\bfR_{diag}}
\newcommand{\Ectrl}{E_{diag}}

\newtheorem{mydef}{Definition}
\begin{document}
%\graphicspath{{/home/jemima/Dropbox/UoR/PhD/Papers/MOpaper/MOFigures/}}
%\graphicspath{{/MOFigures/}}

%PUT THIS IN WHEN USING QJ STYLE FILE
%\runningheads{J. M. Tabeart et al}{Using reconditioned correlated observation error statistics}

\title{The impact of using reconditioned correlated observation error covariance matrices in the Met Office 1D-Var system}

\author{ Jemima M. Tabeart  \thanks{School of Mathematical, Physical and Computational Sciences, University of Reading, UK, NERC National Centre for Earth Observation, ({jemima.tabeart@pgr.reading.ac.uk})},  \and Sarah L. Dance \thanks{School of Mathematical, Physical and Computational Sciences, University of Reading, UK} \and Amos S. Lawless \footnotemark[1], \and Stefano Migliorini \thanks{Met Office, Exeter, UK}, \and Nancy K. Nichols \footnotemark[1], \and Fiona Smith \thanks{Met Office, Exeter, UK, now at Bureau of Meteorology, Hobart, Australia} \and Joanne A. Waller \footnotemark[2]}

%\author{Jemima M. Tabeart \affilnum{1,2}, Sarah L. Dance \affilnum{1}, 
%	Amos S. Lawless \affilnum{1,2}, Stefano Migliorini \affilnum{3}, \\ Nancy K. Nichols\affilnum{1,2}, Fiona Smith \affilnum{3,*} and Joanne~A.~Waller \affilnum{1}}

%\address{\affilnum{1} School of Mathematical, Physical and Computational Sciences, University of Reading, UK \ \break \affilnum{2} National Centre for Earth Observation, Reading, UK
%	\break \affilnum{3} Met Office, Exeter, UK \affilnum{*} Now at Bureau of Meteorology, Hobart, Australia}

%\corraddr{Jemima Tabeart, Department of Mathematics and Statistics, University of Reading, PO Box 220, Reading, RG6 6AX. E-mail: jemima.tabeart@pgr.reading.ac.uk}
\maketitle
\flushleft
\begin{abstract}
Recent developments in numerical weather prediction have led to the use of correlated observation error covariance (OEC) information in data assimilation and forecasting systems. However, diagnosed OEC matrices are {\color{black}often} ill-conditioned and may cause convergence problems  for variational data assimilation procedures. Reconditioning methods are used to improve the conditioning of covariance matrices while retaining correlation information. 
In this paper we study the impact of using the `ridge regression' method of reconditioning to assimilate Infrared Atmospheric Sounding Interferometer (IASI) observations in the Met Office 1D-Var system. This is the first systematic investigation of how changing target condition numbers affects convergence of a 1D-Var routine.  This procedure is used for quality control, and to estimate key variables (skin temperature, cloud top pressure, cloud fraction) that are not analysed by the main 4D-Var data assimilation system. Our new results show that the current (uncorrelated) OEC matrix requires more iterations to reach convergence than any choice of correlated OEC matrix studied. This suggests that using a correlated OEC matrix in the 1D-Var routine would have computational benefits for IASI observations. Using reconditioned correlated OEC matrices also increases the number of observations that pass quality control.
However, the impact on skin temperature, cloud fraction and cloud top pressure is less clear.  As the reconditioning parameter is increased, differences between retrieved variables for correlated OEC matrices and the operational diagonal OEC matrix reduce. %    These retrieval differences are smaller than retrieved standard deviation values for over 75\% of IASI observations. Up to 5\% of retrievals have large differences for alternative choices of the OEC matrix. %: of order 1 for cloud fraction, 20K for skin temperature and 900hPa for cloud top pressure. 
As correlated choices of OEC matrix yield faster convergence, using stricter convergence criteria along with these matrices may increase efficiency and improve quality control. 
\end{abstract}


%	Test test test $t\EintC$, $q\CintE$ $CTP\EintCc$ $CF\CintEc$\\
%Para from \cite{stewart08b} \textit{The 1D-Var assimilation also provides estimates of the atmospheric variables not
%represented in 4D-Var. The control vector in 4D-Var is comprised of a subset of the
%full state vector variables, and those variables, such as skin temperature, which are
%not included are unmodifiable. It is therefore crucial to the success of the assimilation
%that these variables are accurately specified prior to the 4D-Var assimilation.
%For example, radiance observations provide information on all atmospheric variables
%and a poorly specified skin temperature is unchangeable, so therefore the control
%vector variables will be fit incorrectly to the observations. The full state vector
%is used in the 1D-Var assimilation, and the analysis values of those variables not
%present in the control vector are passed to 4D-Var. A schematic of the path of IASI
%observations is shown in Figure 1.}
\section{Introduction}
%\textit{Papers to cite: \cite{bormann03} More theoretical work \cite{waller14b,fowler17,waller16,Hodyss15,Smith17}}\linebreak
% 
In numerical weather prediction (NWP) a data assimilation procedure is used to combine observations of the atmosphere  with a model description of the system in order to obtain initial conditions for forecasts. The contribution of each component is weighted by its respective error statistics.
%Data assimilation is a method used to combine information about a dynamical system from observations and computational model simulations. A well known application of data assimilation is to numerical weather prediction (NWP) where observations of the atmosphere and ocean are combined with a model description of the system, weighted by their respective error statistics. 
In recent years, interest in the understanding and use of correlated observation error statistics has grown (e.g. \citet{janjic17}). %\cite{weston11,stewart09,waller16}. %plus more references 
This increased interest has been motivated by results showing that neglecting correlated observation errors hinders forecasts \citep{rainwater15,stewart08}, and that even including poorly approximated correlation structures is better than using uncorrelated error statistics in the presence of correlated errors \citep{stewart13,healy05}.\linebreak %Theoretical work by \citet{fowler17} has also shown that in the case of complementary observation and background error statistics, analysis variance is minimised and the information content is maximised across a range of scales. \linebreak

Previously, uncorrelated observation error statistics were used for all observations, even when it was known that non-zero error correlations were present. Determining error statistics is a non-trivial problem, as they cannot be observed directly and must be estimated in a statistical sense.  It was also thought that it would not be possible to use correlated observation error covariance (OEC) matrices operationally due to the increased computational cost associated with inverting a dense matrix rather than a diagonal matrix \citep{stewart13}. 
The development of a new method to check error consistency by  \citet{desroziers05} was first applied to explicitly diagnose error correlations using the Met Office system \citep{stewart08b}. Since then, the diagnostic introduced in \citet{desroziers05} (henceforth referred to as DBCP) has been used widely at operational centres \citep{weston11,weston14,stewart14,QJ:QJ3097,bormann11,bormann16,campbell16,Gauthier18,Wang2018}, although uncorrelated OEC matrices are still used operationally for most instruments.
Although much of the initial use of the diagnostic to estimate observation errors focussed on interchannel correlations, this has been extended to spatial correlations \citep{waller14,waller16b,waller16c,cordoba16,Michel18}. 
Theoretical work has also demonstrated how well the diagnostic is expected to perform depending on either the accuracy of the initial choice of background and OEC matrices for the single step \citep{waller16} and the iterative form of the diagnostic \citep{menard16,bathmann18}. The use of the diagnostic in data assimilation schemes using localization has also been considered \citep{waller17}.\linebreak

%\textit{Include a sentence on how good the results of the diagnostic are - physical realism, mathematical form of matrices, conditioning etc. Not symmetric, not positive definite}
The output of the diagnostic cannot be used directly in the assimilation procedure. Diagnosed matrices are asymmetric, and some are not positive definite \citep{stewart14,weston14} and are therefore not valid covariance matrices. Typically, the matrices are symmetrised, and negative and zero eigenvalues are set to be small and positive \citep{weston11}. Additionally, diagnosed OEC matrices are often ill-conditioned. This means that small perturbations to the observations will result in large changes to the analysis, and that iterative methods are likely to converge slowly.  %\citet{waller16} provides theoretical insight into the application of the diagnostic; in particular showing that the iterative version of the diagnostic will converge slowly, or not at all, if the assumed observation and background error covariances are proportional. 
Indeed, the direct use of diagnosed matrices has led to problems with non-convergence of the minimisation of the data assimilation procedure \citep{weston11,weston14}. \citet{weston11} suggested that part of these problems were due to small minimum eigenvalues of the diagnosed OEC matrix, $\bfR$. \linebreak

 %I would put a sentence to link the conditioning and the convergence. I'd add this in before the 'This claim is supported ...'
%It has been shown theoretically, for the case where the observation operator is linear, \citep{haben11c} that the conditioning of variational assimilation problems is related to the convergence of their minimization. %Qn 1: Is this legit?
One way to study the effect of changes to the assimilation system on the convergence of the objective function minimisation is by using the condition number of the Hessian of the variational objective function as a proxy for convergence. This was done in \citet{haben11c} for the case of a linear observation operator.
In \citet{tabeart17a} the minimum eigenvalue of the OEC matrix, $\bfR$, appears in bounds on the condition number of the Hessian of the variational assimilation problem, indicating that this term will also be important for convergence of the objective function minimisation. \linebreak

 %\textit{Is it evident that conditioning and convergence are related, or do I talk about it somewhere else?}%This claim is supported by theoretical work in  \cite{tabeart17a}, where the importance of the minimum eigenvalue of $\bfR$ in terms of the convergence of the variational data assimilation problem has been demonstrated. In \cite{tabeart17a}, bounds on the condition number of the Hessian of the variational assimilation problem, which is linked to the convergence speed of the minimisation, were shown to depend on the minimum eigenvalue of $\bfR$.  %\textit{Some sentence about why this is cool}\linebreak
Increased understanding of how the eigenvalues of $\bfR$ affect the convergence of the data assimilation problem motivated investigation into `reconditioning' methods \citep{weston11,weston14,campbell16,tabeart17a}. These methods increase eigenvalues of the matrix $\bfR$ to improve the conditioning of the OEC matrix, while maintaining much of the existing correlation structure of the diagnosed matrix. 
Two methods are commonly used by NWP centres: `ridge regression' which increases all eigenvalues of $\bfR$ by the same amount, and the `minimum eigenvalue' method which changes only the smallest eigenvalues. These methods were investigated theoretically in \citet{tabeart17c} where it was found that both methods increase standard deviations, and that the ridge regression method strictly reduces all off-diagonal correlations. Both methods were compared in an operational system in \citet{campbell16}, where the sensitivity of forecasts to the choice of method was found to be small, but the ridge regression outperformed the minimum eigenvalue method in terms of convergence. A method similar to the minimum eigenvalue method is used at the European Centre for Medium Range Weather Forecasts (ECMWF) \citep{bormann16}, but will not be discussed further in this paper. \linebreak% Alternative methods of reconditioning are used at other NWP centres in their 4D-Var routines, but these will not be discussed in this paper. For example, the European Centre for Medium Range Weather Forecasts (ECMWF)
%use a method similar to the minimum eigenvalue method discussed in \cite{tabeart17c}; small eigenvalues are set to be equal to a threshold value \citep{bormann16}.

%In this article we investigate the impact of implementing reconditioning methods in the Met Office 1D-Var system.
 The aim of this paper is to investigate the use of the ridge regression method within the Met Office system.
 At the Met Office, in addition to the 4D-Variational data assimilation routine (4D-Var) that is used to produce the initial conditions for weather forecasts, a 1D-Variational data assimilation routine (1D-Var) is used for quality control and pre-processing purposes \citep{Eyre89}. The 1D-Var routine assimilates observations individually, and is used to remove observations that are likely to cause problems with convergence in the 4D-Var routine, as well as to estimate model variables that are not included in the 4D-Var state vector \citep{Pavelin14,Pavelin08}.  After the work of \citet{weston11,weston14}, correlated OEC matrices were introduced in the 4D-Var routine for IASI (Infrared Atmospheric Sounding Instrument) and other hyperspectral IR sounders. However, this was not the case for the 1D-Var routine, where a diagonal OEC matrix continues to be used. Previous work found that diagnosed observation error correlations were small for most channels for the 1D-Var routine \citep{weston11,stewart14} and the proportional increase in computational cost was estimated to be large compared with using correlated OEC matrices in 4D-Var \citep{weston14}. \linebreak

%In this paper we investigate the impact of introducing correlated OEC matrices in the 1D-Var routine by considering how reconditioning. We do this by making use of reconditioning methods to incorporate correlation information in an inexpensive manner. 

In this paper we study how the use of reconditioning methods affects the 1D-Var routine when applied to interchannel OEC matrices for the Infrared Atmospheric Sounding Interferometer (IASI). We examine whether the ridge regression method of reconditioning allows us to include correlated observation error information more efficiently than the diagnosed OEC matrix.  This method of reconditioning is used at the Met Office to recondition OEC matrices that are used in the 4D-Var routine. 
 %set all small eigenvalues to be equal to a threshold value \citep{bormann16}; a method which is related to the minimum eigenvalue method as discussed in \cite{tabeart17c}
  We compare a selection of reconditioned OEC matrices with the current diagonal operational error covariance matrix, and an inflated diagonal OEC matrix. This is the first time that multiple levels of reconditioning have been compared systematically in an operational system. We study the impact of reconditioning in terms of the  computational efficiency as well as the effect on important meteorological variables.\linebreak
  
    In Section \ref{sec:Background} the data assimilation problem is defined and the ridge regression method of reconditioning is introduced. In Section \ref{sec:ExperimentalOverview} we provide an overview of the experimental design. In Sections \ref{sec:Results1DVar} and \ref{sec:Results4DVar} we discuss the impact of changing the OEC matrix  on the 1D-Var procedure, and alterations to the quality control and pre-processing for the 4D-Var routine respectively. 
We find that convergence is improved for any of the choices of reconditioning compared to the current %(at the time of the experiments)
 operational choice of OEC matrix. Additionally, increasing the amount of reconditioning results in faster convergence - which corresponds to theoretical results for the linear variational data assimilation problem in \citet{tabeart17a}. However, the quality control procedure is altered by changing the OEC matrix, with a larger number of observations being accepted for reconditioned correlated OEC matrices compared to the current diagonal choice of OEC matrix. 
 We also find that for most variables, the difference between retrieved values for different choices of OEC matrix are small compared to retrieved standard deviations. However, there are a significant minority of observations for which differences are very large. %We hypothesise that these large discrepancies are likely to be caused by changes to cloud for different choices of OEC matrix.
Finally, in Section \ref{sec:Conclusion} we summarise our results and conclusions.

%\textit{Check that it's ok to cite \cite{weston11} - Met Office technical report}
\section{Variational data assimilation and reconditioning}\label{sec:Background} %??????
%Not sure what needs to go into this section, some things that maybe do need to?
%\sectionoverview{\textit{Much of the background will be the contents of Paper I}}
%\sectionoverview{Data Assimilation overview} 
%\sectionoverview{Reference to Desroziers diagnostic, brief overview}
%\sectionoverview{Results from Paper I - this is where the theoretical motivation for increasing the minimum eigenvalue of the observation error covariance matrix when reconditioning comes from.}
%\sectionoverview{Discuss a bit how other centres treat correlated ob error matrices - \cite{rainwater15}, \cite{bormann16}. (Maybe this should go in introduction)}

 \subsection{Data assimilation}\label{sec:BackgroundVar}
In data assimilation, a weighted combination of observations, $\bfy \in \mathbb{R}^p$, with a background, or `prior', field, $\bfx_b \in \mathbb{R}^n$, is used to obtain the analysis, or posterior, $\bfx_a\in\mathbb{R}^n$. The weights are the respective error statistics of the two components. The matrix $\bfR \in \mathbb{R}^{p\times p}$ is the observation error covariance (OEC) matrix and $\bfB \in \mathbb{R}^{n\times n}$ is the background error covariance matrix. In order to compare observations with the background field, the, possibly non-linear, observation operator $H:\mathbb{R}^n\rightarrow\mathbb{R}^p$ is used to map from state space to observation space. The weighted combination is written in the form of an objective function in terms of $\bfx\in\mathbb{R}^{n}$, the model state vector. In the case of 3D-Var the objective function is given by:
\begin{equation} \label{eq:costfn}
\begin{split}
J(\bfx)=&\frac{1}{2}(\bfx-\bfx_b)^T\textbf{B}^{-1}(\bfx-\bfx_b)\\&+\frac{1}{2}(\bfy-H[\bfx])^T\textbf{R}^{-1}(\bfy-H[\bfx]).
\end{split}
\end{equation}
The value of  $\bfx$ that minimises \eqref{eq:costfn} is given by $\bfx_a$. \linebreak

The first order Hessian, or matrix of second derivatives, of the objective function \eqref{eq:costfn} is given by 
\begin{equation}\label{eq:Hessian}
\nabla^2 J\equiv\bfS = \bfB^{-1}+ \bfH^T\bfR^{-1}\bfH,
\end{equation}
where $\bfH\in \mathbb{R}^{p \times N}$ is the Jacobian of the observation operator, $H[\bfx]$, {\color{black} linearised about the current best estimate of the optimal solution of \eqref{eq:costfn}.} \linebreak%$\bfx_a$. In practice, as the analysis, $\bfx_a$, is unknown, $h$ is linearised about 

 %The conditioning of the Hessian provides us with information about the speed of convergence of the minimisation of the objective function \eqref{eq:costfn} \citep{haben11b}.  
 We now define the condition number of a matrix. Let $\lambda_{max}(\bfS) =\lambda_1(\bfS) \ge \dots \ge \lambda_N(\bfS)=\lambda_{min}(\bfS)$ be the eigenvalues of $\bfS$. We note that this ordering convention will be used for the remainder of the paper. Although covariance matrices are symmetric positive semi-definite by definition, in practice $\bfB$ and $\bfR$ are required to be strictly positive definite in order that they can be inverted in \eqref{eq:costfn}. This means that $\bfS$ is symmetric positive definite, and its condition number is given by

 % As the error covariance matrices $\bfB$ and $\bfR$ are taken to be positive definite for applications, and are symmetric by construction, the condition number of $\bfS$ is given by
 \begin{equation}\label{eq:condS}
 \kappa(\bfS) = \frac{\lambda_1(\bfS)}{\lambda_N(\bfS)}.
 \end{equation}
 
 We note that the minimum possible value of the condition number of any matrix is one. The condition number of the Hessian is of interest because it can be used to study the sensitivity of the solution to small changes in the background or observation data \citep[Sec 2.7]{golub96}.
 %this is not accurate.  Do not solve (1) by CG because it is nonlinear.  Solve by Gauss-Newton and solve the inner linearized problem by CG.  Theory for linear problem is really found in Golub&VanLoan  -  but actually by earlier references.  What does Gill et al refer to?
 As \eqref{eq:costfn} is non-linear, it is solved using a sequence of Gauss-Newton iterations with an inner linearised problem solved using the conjugate gradient method \citep{haben11b}.
  The rate of convergence of the minimisation of the linearised problem by a conjugate gradient function can also be bounded by $\kappa(\bfS)$ \citep{golub96}, although this bound is quite pessimistic. In particular, clustering of eigenvalues can result in much faster convergence than is predicted by $\kappa(\bfS)$ \citep{Nocedal06}. 
  
   %\textit{Otherwise we could get rid of obs to obtain a full rank matrix, practical benefit but also require inverse}
% and as their inverses are required in \eqref{eq:costfn}, $\bfB$ and $\bfR$ are required to be strictly positive definite. 
%This means that $\bfS$ is symmetric positive definite, and its condition number is 

 %The condition number of the Hessian provides an indication of the sensitivity of the solution to small changes in the 
 %We note that the ordering of eigenvalues of matrix $\bfD$ used here: For a matrix $\bfD \in \mathbb{R}^{N \times N}$, let $\lambda_{max}(\bfD) = \lambda_1(\bfD) \ge \lambda_2(\bfD) \ge \dots \ge \lambda_N(\bfD) = \lambda_{min}(\bfD)$.  This ordering convention will be used for the remainder of the paper.
 
 \subsection{Reconditioning: motivation and definition}\label{sec:Recond}
 
 %\subsection{Estimation of observation errors}\label{sec:Desroziers}
 %Many NWP centres use the diagnostic procedure introduced by \cite{desroziers05} to estimate observation error correlations. This diagnostic makes use of observation minus background (O-B) and observation minus analysis (O-A) statistics to calculate the entries of $\bfR$ using 
% \begin{equation}\label{eq:Desroziers}
%\bfR = E\left[\{\bfy-H(\bfx_a)\}\{\bfy-H(\bfx_b)\}^T\right].
% \end{equation}
% The use of the diagnostic relies on two assumptions
% \begin{itemize}
% 	\item Observation and background errors are independent. This assumption is usually satisfied.
% 	\item The observation and background error covariance matrices used to produce the analysis term, $\bfx_a$, in \eqref{eq:Desroziers} are exactly correct. This assumption is violated, as the original $\bfR$ used is uncorrelated, and its errors have been inflated. %re-write this sentence
% 	% \cite{waller16b} \cite{waller16c}
%
% \end{itemize}
In \citet{weston11}, observations from IASI were used at the Met Office for an initial study investigating the feasibility of using correlated observation error matrices in their 4D-Var system. A first guess of the OEC matrix was obtained using the DBCP diagnostic. \linebreak%An overview of the original purpose of this diagnostic as a consistency check and its subsequent use diagnosing correlated error can be found in \citet{waller16}. % However, the results cannot be used directly in an assimilation scheme, for multiple different reasons.\linebreak
  	  	
  	%  	\todo{Para of sentences from other sections which belong here!} This diagnostic has since been trialled at ECMWF \cite{bormann16}, and is now used to provide operational correlated observation error matrices at the Met Office as part of their 4D-Var variational assimilation scheme. 
  	  	
  	 % 	 To achieve this, the Desroziers diagnostic (see Section \ref{sec:Desroziers}) was used to provide a first guess of the observation error covariance matrix. %This diagnostic has since been trialled at ECMWF \cite{bormann16}, and is now used to provide operational correlated observation error matrices at the Met Office as part of their 4D-Var variational assimilation scheme. \linebreak
  	  	%The application of this diagnostic to IASI observations in the Met Office system is introduced in \cite{weston11}. % This work assessed the feasibility and desirability of implementing correlated observation errors in 4D-Var. %Subsequent to further testing, described in \cite{weston14}, this was made operational in 2014. \linebreak
  	  	One problem that was encountered in \citet{weston11} and \citet{weston14} was the ill-conditioning of the matrix resulting from the DBCP diagnostic. The use of an ill-conditioned OEC matrix can result in slower convergence of a variational scheme \citep{weston14,tabeart17a}.  Similar problems were encountered at ECMWF where a degradation in the forecast was seen when the raw output of the DBCP diagnostic was tested \citep{lupu15}.  \citet{weston11} suggested that the convergence problems were caused by very small minimum eigenvalues of the diagnosed observation error covariance matrix.  \linebreak
  	  	
  	  	\citet{tabeart17a} developed bounds for the condition number of the Hessian in terms of its constituent matrices {\color{black} in the case of a linear observation operator}. This provides an indication of the role of each matrix in the conditioning of $\bfS$, and therefore the convergence of the associated minimisation problem. The bound which separates the role of each matrix is given by
  	  	
  	  	\begin{equation}\label{eq:newboundgen}
  	  	\begin{split}
  	  	\max &\Bigg\{\frac{1+\frac{\lambda_{max}(\bfB)}{\lambda_{min}(\bfR)}\lambda_{max}(\bfH\bfH^T)}{\kappa(\bfB)},
  	  	\frac{1+\frac{\lambda_{max}(\bfB)}{\lambda_{max}(\bfR)}\lambda_{max}(\bfH\bfH^T)}{\kappa(\bfB)}, \\
  	  	& \frac{\kappa(\bfB)}{1+\frac{\lambda_{max}(\bfB)}{\lambda_{min}(\bfR)}\lambda_{max}(\bfH\bfH^T)}\Bigg\} \\
  	  &	\le \kappa(\bfS)  
  	  	\le \Big(1+\frac{\lambda_{min}(\bfB)}{\lambda_{min}(\bfR)}\lambda_{max}(\bfH\bfH^T)\Big)\kappa(\bfB).
  	  	\end{split}
  	  	\end{equation}
  	  	These bounds show that the minimum eigenvalue, $\lambda_{min}(\bfR)$, of the OEC matrix is a key term in the upper bound for $\bfS$, meaning that increasing the minimum eigenvalue of $\bfR$ is a reasonable heuristic for reducing the condition number of $\bfS$ and improving the conditioning of the problem \eqref{eq:costfn}. \linebreak
  	  	
  	  	%estriction of \eqref{eq:newboundgen} to the case  where the background errors are homogeneous (i.e. the covariance matrix can be expressed as the product of a scalar variance and a correlation matrix). 
  	  	In the case that the error covariance matrices can be written as the product of a scalar variance with a correlation matrix, e.g. $\bfR = \sigma_o^2\bfD$ and $\bfB = \sigma_b^2\bfC$, and observations are restricted to model variables, we can simplify the bound \eqref{eq:newboundgen} to
  	  	
  	  		\begin{equation} \label{eq:mybound1}
  	  		\begin{split}
  	  		    \max & \Bigg\{\frac{1+\frac{\sigma_b^2}{\sigma_o^2}\frac{\lambda_{max}(\bfC)}{\lambda_{min}(\bfD)}}{\kappa(\bfC)},	\frac{\kappa(\bfC)}{1+\frac{\sigma_b^2}{\sigma_o^2}\frac{\lambda_{max}(\bfC)}{\lambda_{min}(\mathbf{D})}}\Bigg\}\\& \le \kappa(\bfS) \le \Bigg(1+\frac{\sigma_b^2}{\sigma_o^2}\frac{\lambda_{min}(\bfC)}{\lambda_{min}(\mathbf{D})}\Bigg)\kappa(\bfC).
  	  	\end{split}
  	  	\end{equation}
  	  	
  	  	The qualitative conclusions of \cite{tabeart17a} can be summarised as follows.
  	  \begin{itemize}
  	  	\item The minimum eigenvalue of $\bfR$ was shown to be important for determining both the conditioning of the Hessian, and the speed of convergence of a minimisation procedure. This can be seen in \eqref{eq:newboundgen} and \eqref{eq:mybound1}.%This agrees with the findings of \cite{weston14}, where small minimum eigenvalues of the observation error covariance matrix caused convergence problem in an practical setting.
 \item The ratio of the background and observation variances was also shown to be important for conditioning of the Hessian. This can be seen in \eqref{eq:mybound1} explicitly for the case of direct observations where variances are homogeneous for both background and small scale matrices. However, we expect the conclusion to hold more broadly, for example in the case where all standard deviation values corresponding to an OEC matrix were larger than those corresponding to another OEC matrix, then the bounds would be smaller for the first choice of OEC matrix.
 \item  Although  \eqref{eq:newboundgen} and \eqref{eq:mybound1} separate the contribution of each term, {\color{black}numerical experiments revealed that the level of interaction
between observation error and background error statistics depends on the choice of observation network. Examples of observation operators which yield identical bounds for \eqref{eq:newboundgen} but different dependence of $\kappa(\bfS)$ on $\bfB$ and $\bfR$ were found experimentally in \citet{tabeart17a}.}
  	  \end{itemize} 
  	  	
  	  	These conclusions motivate the use of reconditioning methods. In order to make operational implementation of correlated observation error matrices feasible, it is necessary to reduce the impact of the very small eigenvalues of the matrix $\bfR$ by increasing its condition number. To achieve this, different methods of inflation, or reconditioning are used to improve conditioning of correlation matrices for a variety of applications. The ridge regression method is used to recondition OEC matrices at the Met Office \citep{weston11,weston14}, and hence will be the reconditioning method that is considered in the remainder of this paper.  The ridge regression method adds a scalar multiple of the identity to $\bfR$ to obtain the reconditioned matrix $\bfR_{RR}$. This scalar, $\delta$, is chosen such that $\kappa(\bfR_{RR})=\kappa_{max}$, a user-specified condition number. The method for calculating $\delta$ for a given choice of $\kappa_{max}$ was formally defined in \citet{tabeart17c} as follows:  	  	
  	  	
\begin{mydef}
	\textbf{Ridge regression reconditioning constant, $\delta$ \citep{tabeart17c}}\\
%	\flushleft
%The ridge regression method adds a scalar multiple of the identity to $\bfR$ to obtain the reconditioned matrix $\bfR_{RR}$. The scalar $\delta$ is set using the following method.
%\begin{itemize} %Layout of this - do we want it all on one line? Could we italicise and centre?
	 Define $\delta = ({\lambda_{max}(\bfR) - \lambda_{min}(\bfR)\kappa_{max}})/({\kappa_{max}-1}) $. \\	 
	 Set $\Rp = \bfR+\delta\textbf{I}$
%\end{itemize}
\end{mydef}

%Using this method shifts all of the eigenvalues in the same way.
We note that this choice of $\delta$ yields $\kappa(\Rp)=\kappa_{max}$. %This method is similar to Steinian linear shrinkage \cite{Ledoit04}, which produces a well conditioned estimator to a covariance matrix via an optimal linear combination of the covariance matrix with the identity matrix. The ridge regression method is used at the Met Office \cite{weston14}.
%In this method we increment the diagonal elements of $\bfR$ by $\delta>0$ chosen such that the condition number of $\Rp$ is equal to $\kappa_{max}$. We calculate $\delta$ as follows:
%\begin{equation}
%\delta = \frac{\lambda_{max}(\Rp) - \lambda_{min}(\Rp)\kappa_{max}}{\kappa_{max}-1} .
%\end{equation}
%Using this method shifts all of the eigenvalues in the same way. 
Mathematical theory describing the effect of this reconditioning method on the correlations and variances of any covariance matrix was developed in \citet{tabeart17c}, which showed that the ridge regression method increases error variances for all observations, and decreases all off-diagonal correlations. {\color{black}In this paper we investigate whether the qualitative conclusions from \citet{tabeart17a} hold in the case of a non-linear observation operator, and we study the impact of reconditioning methods in an operational system.}

\section{Experimental Overview}\label{sec:ExperimentalOverview}

%\sectionoverview{Overview of Met Office system (6.1 in project)}
%\sectionoverview{Description of choices of $\bfR$}
%\linebreak
%\sectionoverview{System parameter choices used for experiments}
%\sectionoverview{Description of choice of data to look at - what are our experiments and what are we looking for}

\subsection{Met Office System} \label{sec:MOsystem}
%\todo{Do we want to combine these two sections? See how long this one is when written up}
%Motivation/intro: For this we will be looking at data from one type of obs. This is IASI - explain a bit, why did we pick it, what does it measure. (Current first para)
%\begin{itemize}
%	\item What observations are we using and why (briefly)
%	\subitem Channels sensitive to water vapour - could make better use of these channels with the introduction of correlations.
%	\item What is the 1D-Var routine used for, and why are we interested in it
%	\subitem Quality control, observations that are fixed and not assimilated by 4D-Var.
%	\subitem Currently uncorrelated choice of $\bfR$, even though 4D-Var routine has correlated
%	\subitem Could it be beneficial to move to correlated $\bfR$?
%\end{itemize}
The experiments carried out in this manuscript will use observations from the IASI instrument on the EUMETSAT MetOp constellation. IASI is an infrared Fourier transform spectrometer, and measures infrared radiation emissions from the atmosphere and surface of the earth \citep{chalon01}. {\color{black} We note that the observation operator for this instrument, a radiative transfer model, is highly non-linear so the conclusions from \citet{tabeart17a} will not necessarily apply to this problem.}
%A single observation of emitted radiation at one location produces measurements across the infrared spectra
The infrared spectrum is split into channels corresponding to different wavelengths; this means that an observation at a single location will provide information for up to 8641 channels.
An early use of the DBCP diagnostic focused on observations from IASI implemented in the Met Office system \citep{stewart08b}.
%The first investigation at the UK Met Office into the introduction of correlated small scale matrices made use of diagnosed IASI errors \cite{stewart08}. 
Much of the subsequent research on correlated observation error uses IASI observations \citep{weston11,weston14,stewart14,bormann16}. %, partly due to the fact that the existence of correlated observation errors has been shown theoretically for this instrument \cite{stewart09}. 
In particular IASI has channels that are sensitive to water vapour which have been found to have errors with large correlations \citep{stewart14,weston14,bormann16}. \linebreak%The use of correlated small scale matrices in the data assimilation procedure may allow better use of these channels.

One attraction of IASI, and other hyperspectral instruments, is the large number of available channels, which provides high vertical resolution. However, using all of these channels is not feasible in current operational NWP systems for reasons including computational expense, and not requiring too many observations of a similar type.   Additionally, when IASI was first used, there was a reluctance to include correlated observation errors so an effort was made to choose channels that are spectrally different and hence less likely to have correlated errors \citep{stewart14}. This means that of the 8461 available channels, only a few hundred are used at most NWP centres \citep{stewart14}. At the {\color{black}time of the experiments, the Met Office stored a subset of $314$ channels with a maximum of $137$ being used in the 4D-Var system. A list of these channels is given by \citet[Appendix A]{stewart09}.} 
As there is a large degree of redundancy between channels \citep{Collard10}, directly assimilating a larger number of channels is likely to make the conditioning of the OEC matrix worse. This has motivated alternative approaches such as principal component compression \citep{Collard10} and the use of transformed retrievals \citep{prates16}, which will not be considered in this work.\linebreak

 A larger number of channels is used for the 1D-Var assimilation than for the Met Office 4D-Var assimilation; standard deviation values for these channels are filled in from the current operational (diagonal) OEC matrix. We chose to focus on the channels used in the 4D-Var system in order to be consistent between both assimilation systems.
We also note that not all channels are used for each assimilation; for example, some channels are not used in the presence of cloud. In this case, rows and columns corresponding to channels that are affected by cloud are deleted from the OEC matrix. As the submatrix chosen from the full OEC matrix used could change at each observation time, there may be a difference in the condition number of the OEC matrix used in practice and the OEC matrices presented in this work.
%presented in this work may differ from the condition number of the OEC matrices that used
%This means that the condition number may be different for the OEC matrices that are used in practice. 
However, the Cauchy interlacing theorem \citep[Lemma 8.4.4]{Bernstein} states that  the condition number will not be increased by deleting rows and columns of a symmetric positive definite matrix. This means that the values given here are upper bounds for $\kappa(\bfS)$ even if the quality control procedure excludes some channels. \linebreak

We test the impact of using correlated OEC matrices in the Met Office 1D-Var system and consider the effect of using the ridge regression method of reconditioning with different choices of target condition number. At the Met Office, 1D-Var is run prior to every 4D-Var assimilation procedure, meaning that retaining current computational efficiency and speed of convergence is desirable. We note that a single IASI observation consists of brightness temperature values for each of the channels that are used in the assimilation. 
A 1D-Var procedure takes observations separately at each location to retrieve variables such as temperature and humidity over a 1D column of the atmosphere. This procedure is much cheaper and more parallelisable than a 4D-Var algorithm.\linebreak

The 1D-Var routine performs two main functions: %(1) quality control and (2) estimation of values for certain variables that do not form part of the 4D-Var state vector:
\begin{enumerate}
	\item Quality control (QC): %There are multiple aspects to the QC procedure which depend on the choice of OEC matrix. Firstly a number of cloud detection tests are conducted to remove observations that are affected by cloud \cite{hilton09}. One test of these tests, which is described in \cite{english99}, uses an objective function which includes the inverse OEC matrix. Therefore changing the OEC matrix will alter the cloud detection procedure, and which observations are rejected due to the presence of cloud. \linebreak%We also note that observations from IASI are compared with collocated observations from AMSU, another satellite instrument, and rejected if they exceed a threshold of 5K. This means that physically infeasible observations are rejected. \linebreak 
	Observations that require more than 10 iterations for the 1D-Var minimisation to reach convergence are not passed to the 4D-Var routine. This is because it is assumed that observations for which the retrieval procedure takes too long to converge for the 1D-Var minimisation will also result in slow convergence for a 4D-Var minimisation. {\color{black} The convergence criteria is based on the value of the cost function and normalised gradient \citep{Pavelin08}.} Changing the OEC matrix will alter the speed of convergence of 1D-Var, and hence affect which observations are accepted. %{\color{black}Channels sensitive to temperature below the cloud top are excluded via a method described in \citet{Pavelin08}.}  \linebreak
	\item Estimation of values for certain variables that are not included in the 4D-Var state vector:  values for skin temperature, cloud fraction, cloud top pressure and emissivity over land are fixed by the 1D-Var procedure. Altering the OEC matrix will change retrieved values for these variables.
\end{enumerate}

Changing the OEC matrix is therefore likely to have two main effects on results of the 1D-Var procedure: changing the observations that are accepted by the quality control, and changing the values of those variables not included in the 4D-Var state vector.  Skin temperature (ST), cloud fraction (CF) and cloud top pressure (CTP) are retrieved as scalar values at each observation location. In contrast, surface emissivity is retrieved as a spectrum, which is represented as a set of leading principal components \citep{Pavelin14}. As we expect the interactions between the choice of $\bfR$ and the retrieved values to be complex, in this work we only consider the effect of changing $\bfR$ on the three scalar variables: skin temperature, cloud top pressure and cloud fraction.

\subsection{Experimental Design} \label{sec:experimentaldesign}
%\begin{itemize}
%	\item What are we interested in? How do outputs of 1D-Var change as we change $\bfR$? Split this into two parts: analyse effect on 1D-Var (speed of convergence, retrievals) and in effect on variables that are passed to 4D-Var.
%	\item What data are we using? (Does this belong in previous section?)
%	\subitem Carried out trials over a range of dates, but all plots here show 16th June 2016 0000Z. Results across different times of day/year were similar and not shown.
%	\subitem Observations from Meteorological Data Base, with background from Unified Model background files for global configuration.
%	\subitem Fixed operational choice of $\bfB$.
%	\item What are we changing?
%	\subitem We consider 7 choices of $\bfR$. List in a less dense way than currently! Keep table 1.
%\end{itemize}
%Set-up general
We now describe the experimental framework and key areas of interest that will be investigated in Sections \ref{sec:Results1DVar} and \ref{sec:Results4DVar}. We use the operational Met Office 1D-Var framework at the time of the experiments (July 2016), and consider how the results change for different choices of OEC matrix. Background profiles are obtained from the Unified Model (UM) background files for the corresponding configuration.  A number of different times and dates for the six months between December 2015 and June 2016 were considered, but as results were similar across all trials we only present results from experiments for 16th June 2016 0000Z. \linebreak

%Describe B matrix

%Set-up R matrices
%We now describe the choice of OEC matrices that will be used.  
The correlated choices of $\bfR$ are calculated using the method introduced in \citet{weston14}; applying the ridge regression method of reconditioning to the diagnosed matrix for a variety of choices of $\kappa_{max}$.
The matrices estimated by the DBCP diagnostic depend considerably on the choice of background and observation error matrices. For all OEC matrices produced, the same $4$ days of IASI and background NWP data (03/12/15-06/12/15), were used as input data. We note that the estimated OEC matrix was obtained using background and OEC matrices  from the 4D-Var assimilation routine rather than the 1D-Var routine. 
{\color{black} Although this is not theoretically consistent with the smaller error correlations that have been estimated for the 1D-Var problem in previous studies \citep{stewart14,weston11}, the use of 4D-Var error statistics allows us to better understand the impact that our changes are likely to have on 4D-Var. We are using 1D-Var as a pre-processing step for 4D-Var to remove observations that are likely to cause convergence issues in the main assimilation algorithm.}  \linebreak{}%Additionally, a major aim of this paper is to consider the impact of reconditioning methods on an operational system, and our conclusions will still hold in the case that an `incorrect' matrix is used. This, combined with the fact that some correlation information is better than none \citep{healy05}, motivates our decision to study the OEC matrix arising from 4D-Var.}   \linebreak%{\color{red} Need to justify this decision.}  
 %This means that for all tests the underlying raw Desroziers matrix was the same. We note that these raw diagnosed errors come from application of the diagnostic to the background and OECs of the 4D-Var routine, not the 1D-Var routine. For a full description of this procedure see \cite{stewart14}. 
 \begin{figure*}
    \centering
    \includegraphics[width=0.85\linewidth]{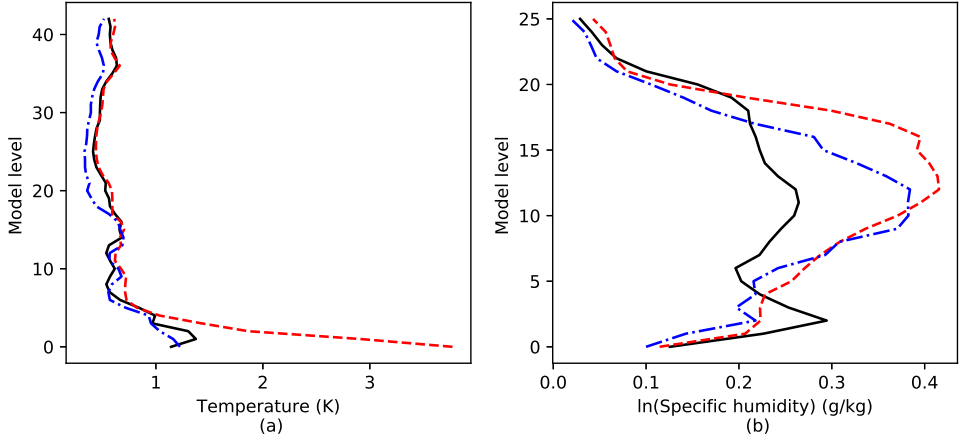}
    \caption{Standard deviation values for the operational background error covariance matrices, $\bfB$, for the northern hemisphere (solid line), tropics (dot-dashed line) and southern hemisphere (dashed line) for temperature (a) and ln(specific humidity) (b).}
    \label{fig:Bsd}
\end{figure*}
\begin{figure*}
    \centering
    \includegraphics[trim=30mm 0mm 30mm 0mm,clip,width=1\linewidth]{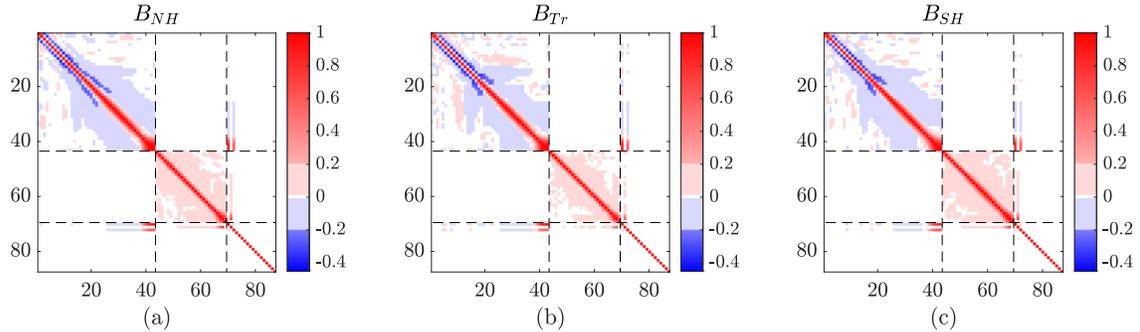}
    \caption{Correlation matrices for the operational background error covariance matrices, $\bfB$, for the northern hemisphere (a), tropics (b) and southern hemisphere (c). Dashed vertical and horizontal lines separate inter and cross correlations between temperature, ln(specific humidity) and other variables (from left to right).  ST is variable 72, CTP is 74 and CF is 75. }
    \label{fig:Bcorr}
\end{figure*}

\begin{table*}
{\color{black}\centering
\begin{tabular}{| l | c | c| c | c | c| }
\hline		
Variable  & CF & CTP (hPa) & ST (NH) (K) & ST (Tr) (K) & ST (SH) (K) \\	
\hline
Standard deviation & 1 & 1000 & 2.24 & 1.92 & 2.02\\
\hline
%\centering
\end{tabular}

\caption{Background standard deviation values for variables not included in the 4D-Var state vector. }
\label{tab:Bsd}}
\end{table*}

{\color{black}We use the operational background error covariance matrix, $\bfB$, at the time of the experiments. This consists of three different choices of $\bfB$ for the northern hemisphere (30N:90N), the tropics (30S:30N) and the southern hemisphere (90S:30S). 
Figure \ref{fig:Bsd} shows background standard deviation (BSD) values for temperature and humidity variables, and Table \ref{tab:Bsd} gives BSD values for CF, CTP and ST for each of the choices of $\bfB$. We note that standard deviations for cloud variables are assumed to be very large so that the background is ignored for these variables \citep{Pavelin08}.  In Sections \ref{sec:Results1DVar} and \ref{sec:Results4DVar} we will compare the standard deviations from the background error covariance matrix against retrieved standard deviations for the observations as well as differences between observations for different choices of $\bfR$. 
Figure \ref{fig:Bcorr} shows that correlations corresponding to the three choices of $\bfB$ are qualitatively very similar.
Cross-correlations between variables are quite weak, with no correlations between temperature and specific humidity. 
Most correlations larger than 0.2 occur for adjacent model levels for temperature and specific humidity.
Correlations greater than 0.2 also occur between surface temperature and ST and temperature for larger model level numbers, and surface specific humidity and specific humidity at larger model level numbers. CTP and CF are uncorrelated with all other variables. }\linebreak
% Values for CTP = 1000, CF = 1, ST = 2.24194046754146,1.91966281935136,2.01822274538763 (NH,Tr,SH)

  %CHANNEL SELECTION

We apply the DBCP diagnostic to the  subset of $137$ channels that are assimilated in the 4D-Var routine. The 1D-Var routine uses additional channels \citep{hilton09}, with a total of 183 channels being assimilated. Observation errors for these additional channels are assumed to be uncorrelated, and filled in with values from the diagonal error covariance matrix $\bfR_{diag}$. %We use $\bfR \in \mathbb{R}^{137\times 137}$ to denote the covariance matrix corresponding to the subset of channels, and $\bfR^{334}\in \mathbb{R}^{334\times334}$ to denote the covariance matrix corresponding to the full set of channels. 
If additional channels are included in future versions of the operational system, it would be advisable to recompute the DBCP diagnostic applied to all channels. \linebreak
The seven different choices of the matrix $\bfR$ that were tested are now listed: %We note that all of these are of dimension $334 \times 334$: 
\begin{itemize}
	\item  $\bfR_{infl}$ which is an inflated diagonal matrix. %We note that this is the same as the operational $\bfR$ for \citet{weston11} for the 4D-Var algorithm. 
	This matrix was used prior to the introduction of correlated observation error in the 4D-Var assimilation scheme \citet{weston14}. In particular variances are inflated to account for the fact that the assumption of uncorrelated errors is incorrect. The standard deviations (square root of the diagonal entries of $\bfR_{infl}$) are shown in \citet[Figure 1]{weston14} by the black dashed line. The largest value entry of $\bfR_{infl}$ is $16$, and the smallest entry is $0.25$. The construction of $\bfR_{infl}$ is described in \citet{hilton09}. %This is the matrix that is used to fill in the missing channels when applying the Desroziers diagnostic.
	\item $\Rctrl$, the current operational matrix for 1D-Var retrievals, which is diagonal.  The standard deviations are calculated as instrument noise plus $0.2K$ forward-model noise \citep{Collard07}. The variances of $\Rctrl$ are shown in \citet[Figure 7]{stewart14} by the red line. The variances are much smaller than for $\bfR_{infl}$; for the first $120$ channels, the diagonal elements of $\Rctrl$ are all less than $ 0.27$ and the largest value of $\Rctrl$ is given by $0.49$.
	\item $\bfR_{est}$, the symmetrised raw output of the code that produces the DBCP diagnostic. This is computed by $\bfR_{est} = \frac{1}{2}(\bfR_{DBCP}+\bfR_{DBCP}^T)$, where $\bfR_{DBCP} \in \mathbb{R}^{137\times 137}$ is the output of the DBCP diagnostic.% with filled in missing channels from the diagonal matrix $\bfR_{infl}$.
	\item Reconditioned versions of $\bfR_{est}$ so that the correlated submatrix has a condition number of $1500$, $1000$, $500$ and $67$, referred to respectively as $\bfR_{1500}, \bfR_{1000}, \bfR_{500}$ and $\bfR_{67}$. 
	\end{itemize}
We refer to the experiments using each choice of OEC matrix as $E$ with subscript corresponding to that of the OEC matrix (i.e. $\Ectrl, E_{est}, E_{1500}E_{1000}, E_{500}, E_{67}$ and $E_{infl}$). \linebreak%We also collectively refer to $E_{est}, E_{1500}, E_{1000}, E_{500}, E_{67}$ and $E_{infl}$ as $E_{exp}$

\begin{table*}
\centering
\begin{tabular}{| l | c | c| c | c | c|c|c| }
\hline		
Experiment name & $\Ectrl$ & $E_{est}$ & $E_{1500}$ & $E_{1000}$ & $E_{500} $ &$E_{67}$& $E_{infl}$ \\	
\hline
Choice of $\bfR$ & $\Rctrl$ & $\bfR_{est}$ & $\bfR_{1500}$ & $\bfR_{1000}$ & $\bfR_{500} $ &$\bfR_{67}$ & $\bfR_{infl}$\\
\hline
$\lambda_{min}(\bfR)$  & 0.025 & 0.00362 & 0.00482 & 0.007244 & 0.0145 & 0.1010 & 0.0625 \\
\hline
%$\kappa(\bfR^{334})$ (full-matrix)& 1600 & 2844 & 27,703 & 20,712 & 13,804 & 6895 & 1600 \\
%\hline  (137 correlated channels only)
$\kappa(\bfR)$  & 9.263 &2730 & 1500 & 1000 & 500 & 67& 64 \\
\hline
%\centering
\end{tabular}

\caption{Minimum eigenvalues and condition number of $\bfR$ for each experiment.}
\label{tab:mineig}
\end{table*}
%How much detail am I going to go into about this?

%RTTOV
%IASI - what does it measure, how do we assimilate in practice?
%Overview of previous work in this area

%\textit{What experiments/outputs are we going to consider?}
Details of the conditioning, and minimum eigenvalues of each of the choices of $\bfR$ can be found in Table \ref{tab:mineig}.  We see that for the non-diagonal matrices, as we decrease the target condition number, we increase the minimum eigenvalue of $\bfR$. This agrees with the theoretical results of \citet{tabeart17c}. We also see that of the two diagonal choices of $\bfR$, $\bfR_{infl}$ has the larger value of $\lambda_{min}(\bfR)$, suggesting that we might expect better convergence compared to $\Rctrl$.  We also notice that the largest value of $\lambda_{min}(\bfR)$ occurs for $\bfR_{67}$. It will be of interest to consider whether the introduction of correlations has more effect on convergence and conditioning than the value of $\lambda_{min}(\bfR)$. We note that the inclusion of $46$ extra channels  in the 1D-Var algorithm, in addition to the $137$ channels used in the 4D-Var algorithm, could change the condition numbers presented in Table \ref{tab:mineig} by the introduction of very small or very large eigenvalues. \linebreak%In practice this would mean that beyond a certain point, increasing the reconditioning parameter $\delta$ may not alter the condition number.  %Table \ref{tab:mineig} also shows the condition numbers of the full matrices in comparison with the condition number of the subset of 137 correlated channels. We note that the condition numbers for $\bfR^{334}$ are much larger than for the corresponding choice of $\bfR$. However, the full matrix is not used in practice, so these large condition numbers will not lead to convergence problems. We also see that although  $\kappa(\bfR_{67}^{334})=\kappa(\bfR_{infl}^{334})$, the condition numbers of the correlated sub-matrices are different. This is because the smallest eigenvalue of the uncorrelated sub-matrix is smaller than $\lambda_{N}(\bfR)$ for these choices of OEC.%%Although the conditioning of $\bfR^{334}$ is very large for all choices of observation error covariance, as the full matrix is not used in and so will not lead to convergence problems.

Our numerical experiments will be broadly split into two groups. Firstly we will consider the effect of changing the OEC matrix, $\bfR$, on the 1D-Var procedure itself in Section \ref{sec:Results1DVar}. This includes the impact on retrieved values and the convergence of the 1D-Var assimilation. Secondly, in Section \ref{sec:Results4DVar}, we will consider the impact of these changes on the 4D-Var procedure, by looking at how the number of accepted observations varies, and how the retrieved values of skin temperature, cloud top pressure, and cloud fraction retrievals are altered.

\section{Impact on Met Office 1D-Var routine}\label{sec:Results1DVar}

%In the previous section theoretical bounds on the condition number of the Hessian of the variational data assimilation problem were established for the unpreconditioned case where both $\bfB$ and $\bfR$ are correlated. In particular, bounds which separated the contribution of each constituent matrix were developed. Numerical experiments that tested these bounds in an idealised linear context were also presented. For many instruments, operational observation operators are highly non-linear \cite{stewart09}. It is therefore of interest to test our theoretical conclusions in an operational setting to see if they are still applicable in the non-linear case. \linebreak

In this section we consider the impact of changing the OEC matrix used in the Met Office 1D-Var system on the conditioning of the Hessian and on individual retrievals of temperature and humidity.
In particular, the conditioning of the Hessian is important in terms of speed of convergence of the minimisation procedure. We recall (Section \ref{sec:MOsystem}) that in 1D-Var information for each observation location is assimilated separately. Here a single observation corresponds to information from a column of IASI channels valid at one location.  {\color{black} This corresponds to 97330 observations over the $4$ days of data discussed in Section \ref{sec:experimentaldesign} with objective functions that  converge in 10 or fewer iterations for all choices of $E_{exp}$. For much of the discussion that follows we will consider statistics of this set of 97330 observations to understand how changing the OEC matrix affects 1D-Var for IASI observations.}  %DO we want to justify our interest in temp and humidity retrievals? Can't think of anything better than it tells us what impact correlated R has, which is pretty obvious already.

%This section describes results from experiments carried out using the Met Office system  \todo{Correct references/re-write for current paper}. The framework used will be introduced along with some concerns that are specific to operational scale problems. Results from tests carried out using 1D-Var in the Met Office global model will be discussed. %These experiments will principally focus on testing the conclusions from Section \ref{sec:NewWork} in the operational setting, however some topics not fully investigated in previous sections will be explored here. % Variables of interest include the number of iterations required for convergence of the 1D-Var retrieval, and results of these retrievals. %As well as investigating the impact of different choices of $\bfR$ on the condition number of the Hessian of the cost function, it is also of interest to see how changing $\bfR$ changes the minimal value of the cost function. The results of these experiments will be discussed and related to the theory developed in Section~\ref{sec:NewWork}. 
%Many of the ideas for these tests follow on from work in \cite{stewart09}, \cite{haben11c} and in particular \cite{weston14}.

%Say something about previous work done in this area
%The main result from the previous section that we wish to test is the bound from Corollary \ref{lemma:factorisedprecond}.\linebreak

%In this chapter we will provide an overview of the experiments on the Met Office system, and their results.

\subsection{Influence of observation error covariance matrix on convergence and conditioning of the 1D-Var routine} \label{sec:MOHessian}
We begin by investigating explicitly the effect of changing the OEC matrix, $\bfR$, on the 1D-Var routine. We consider two variables: the number of iterations required for convergence for the minimisation routine and the condition number of the Hessian of the 1D-Var cost function.\linebreak% \citet{tabeart17a} showed that the the choice of OEC matrix plays an important role in the conditioning and convergence of a simple linear system, so we now investigate whether the conclusions presented in that work hold in a complex non-linear system. \linebreak

Firstly we consider the number of iterations required for the minimisation of the 1D-Var cost function to reach convergence for each assimilated observation. For NWP centres, this is a variable of significant interest, as the extra expense of introducing correlated error predominantly comes from the increase in the number of iterations needed before convergence in the case of interchannel errors \citep{weston11}. We note that this may not be the case for other types of error correlation such as spatial and temporal correlations (where the computation of matrix-vector products may require additional communication between processors \citep{simonin18}). %At the Met Office, the 1D-Var minimization allows a maximum of $10$ iterations in the current setting, although for the majority of observations, convergence of the minimization of the  cost function requires fewer than $3$ steps in the operational (uncorrelated) case. \linebreak
The minimisation is deemed to have converged when the absolute value of the difference between each component of two successive estimates of the state vector is smaller than $0.4\sigma_{\bfB}$, where $\sigma_{\bfB}$ is the vector whose components are the background error variances for each retrieved variable. Values deemed to be unphysical, such as temperature components falling out of the range $70K-340K$, are discarded.  \linebreak%WOULD be good to get a reference for the cost of running OPS at some point
%The stopping criteria used will be discussed here when I have confirmed which of two methods is used. \linebreak
\begin{figure*}
	\centering
%	\begin{subfigure}{0.38\textwidth}
		\centering
		\includegraphics[trim= 0mm 3mm 0mm 1mm,clip,width=0.75\linewidth]{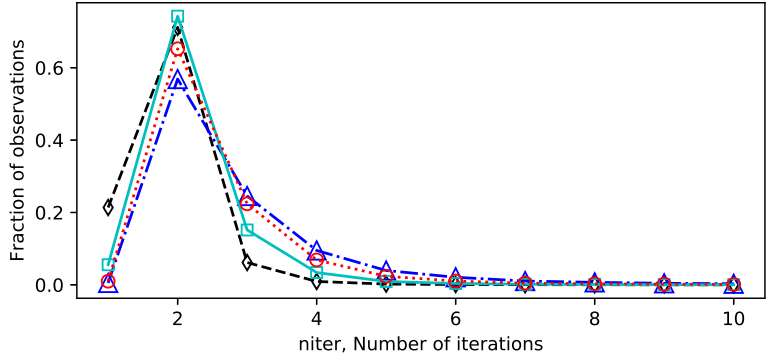}

	%\end{subfigure}	
%	\begin{subfigure}{0.38\textwidth}
%		\centering
%		\includegraphics[width=\linewidth]{Hesscondnooutliers}
%		\caption{Cumulative histogram of iterations}
%		\label{fig:Hesscondnooutliers}
%	\end{subfigure}		
	\caption[Number of iterations]{Number of iterations required for convergence of the minimization of the 1D-Var cost function as a fraction of the total number of observations common to all choices of $\bfR$. Symbols correspond to: $\Rctrl$ ($\triangle$), $\bfR_{est}$ ($\circ$), $\bfR_{67}$ ($\square$) and $\bfR_{infl}$ ($\diamond$). 			\label{fig:niter}}
%	\label{fig:niterboth}
\end{figure*}

\begin{table*}
	\centering
	\begin{tabular}{ l  c c c  c c c c  }
		\hline
		 & $ \Ectrl$ & $E_{est}$ & $E_{1500}$ & $E_{1000}$ & $E_{500}$ & $E_{67}$ &$ E_{infl}$ \\
		\hline
\small{}&\small{}&\small{}&\small{}&\small{}&\small{}&\small{}&\small{} \\
		$\max \kappa(\bfS)$ & $3.01\times 10^{12}$ & $7.546\times 10^{11}$ & $7.469\times 10^{11}$ & $7.30\times 10^{11}$ & $7.02\times 10^{11}$ & $3.71\times 10^{11}$ & $1.74\times10^{11}$ \\
		mean $\kappa(\bfS)$ &$2.78\times 10^{10}$ & $ 6.71\times 10^9$&$ 6.62\times 10^9$& $6.43\times10^9$& $6.00\times10^9$&$ 4.01\times10^9$ &$2.83\times10^9$ \\
		median $\kappa(\bfS)$ &$2.09\times10^8$&$ 1.31\times10^8$&$ 1.32\times10^8$&$ 1.33\times10^8$&$ 1.37\times10^8$&$ 1.78\times10^8$&$ 2.89\times10^8$\\
		\hline
	\end{tabular}
	\caption{Maximum, mean and median values of $\kappa(\bfS)$ for $E_{diag}$ and experiments.}
	\label{table:condHess}
	
\end{table*}

For each observation, we store the number of iterations required for the corresponding 1D-Var objective function to converge, $niter$. Figure \ref{fig:niter} shows the fraction of observations that have objective functions that converge in $niter$ iterations for four choices of $\bfR$. We note that the behaviour for the other correlated experiments is similar to the behaviour for $E_{est}$ and hence only the distributions for $E_{est}$ and $E_{67}$ are shown. %Values are normalised so that they sum to one across each choice of $\bfR$. 
 We see that for all experiments $niter = 2$  is the modal class and contains over $50\%$ of the observations. We begin by considering experiments corresponding to correlated choices of the matrix $\bfR$. Our results show that as the minimum eigenvalue of the matrix $\bfR$ increases, there is a decrease in the required number of iterations. %The distributions for experiments $E_{est}$, % ($\circ$), 
%$E_{1500}$, % ($\triangledown$), 
%$E_{1000}$, %($\triangle$) 
%and $E_{500}$ %($+$)
 %are very similar, hence only the distribution for $E_{est}$ is shown.
 %Generally, the mode of the probability density function (pdf) shifts to the left (i.e. there is a decrease in the required number of iterations) as the minimum eigenvalue of the matrix $\bfR$ increases.  This shift to the left is seen more clearly for $E_{67}$ which has the highest proportion of observations corresponding to $niter = 2$ iterations, and noticeably fewer for $niter=3$ and $niter=4$ than any other choice of correlated OEC matrix.
 %A plot of the cumulative probability distribution for the correlated experiments (not shown here), shows that the fraction of observations which correspond to $niter\le k$ increases with $\lambda_{min}(\bfR)$ for all choices of $1\le k\le 10$. 
 This agrees with the theoretical conclusions of \citet{tabeart17a}.    
However, the overall effect of reconditioning on convergence speed is less for 1D-Var than was observed in the case of 3D-Var or 4D-Var as described in \citet{weston11}.  It is likely that this is because the average number of iterations is greater in 3D and 4D-Var, and the maximum permitted number of iterations is much larger than the $10$ allowed for the 1D-Var minimisation. \linebreak

 We now consider the two diagonal choices of OEC matrix, $\bfR_{infl}$ and $\Rctrl$. 
The distribution corresponding to $\Ectrl$ is more heavily weighted towards a higher number of iterations than any of the correlated cases. This is not what we might expect from an uncorrelated choice of OEC matrix, particularly as it is well-conditioned compared to most other choices of OEC matrix.
In particular, $\lambda_{min}(\Rctrl)$ is greater than the minimum eigenvalue for all choices of correlated OEC matrix apart from $\bfR_{67}$ (see Table \ref{tab:mineig}).  
In contrast, for the experiment $E_{infl}$ %we see that a larger proportion of observations correspond to objective functions that converge in a single iteration. For $niter>2$ the proportion of observations for $E_{infl}$ is smaller than the other choices of experiment. This means that
convergence is faster than for any of the other experiments.\linebreak

{\color{black} As we noted in Section \ref{sec:Recond}, the minimum eigenvalue of the matrix $\bfR$ is not the only important property for determining the speed of convergence.  
%The difference in convergence between $\Ectrl$ and $E_{infl}$ can be explained by considering T
The distribution of standard deviations for $\Ectrl$ and $E_{infl}$ is shown in Figure 1 of \citet{weston14}. As the standard deviations for $\bfR_{infl}$ are much larger than the standard deviations for any other choice of $\bfR$, the ratio of background variance to observation variance will be smaller for $E_{infl}$ than other experiments, resulting in smaller condition numbers of the Hessian and hence faster convergence of the 1D-Var minimisation. We recall from Section \ref{sec:Recond} that the ratio of background to observation error variances appears in the bounds on the condition number of the Hessian given by \eqref{eq:mybound1} in \citet{tabeart17a} and similar bounds in \cite{haben11c}. It is clear from these bounds that decreasing the observation error variance will increase the value of the bounds.
We can therefore explain %However, variances for $\Rctrl$ are similar in size to variances for $\bfR_{est}$, so this does not explain 
the worse convergence seen for $\Ectrl$ by considering channels 107-121 and 128-137, where variances for $\Rctrl$ are smaller than the variances for correlated choices of $\bfR$.} These channels are sensitive to water vapour, and also correspond to the strongest positive correlations in $\bfR_{est}$. Typically, inflation is used when correlated errors are not accounted for; here we have the opposite effect with smaller variances for uncorrelated $\Rctrl$. 
 In terms of the minimisation of the 1D-Var objective function, this means that $\Ectrl$ is pulling much closer to observations for those channels than any of the correlated experiments. This makes it harder to find a solution, resulting in slower convergence.  \linebreak% \halfblankline% As the estimated variables will be very different depending on how close the data assimilation algorithm fits different channels, 
We now consider how the condition number of the Hessian of the 1D-Var cost function, $\kappa(\bfS)$, changes with the experiment $E$. From theoretical results developed in \citet{tabeart17a}, in particular the result of Corollary~1, we expect $\kappa(\bfS)$ to decrease as $\lambda_{min}(\bfR)$ increases. The minimum eigenvalues for each choice of OEC matrix, $\bfR$, discussed here can be seen in Table \ref{tab:mineig}.  The condition number of $\bfS$ is computed separately for each objective function. We can therefore consider the maximum, mean and median value of $\kappa(\bfS)$ over the 97330 observations for each experiment. This information is shown in Table~\ref{table:condHess}. As discussed in Section \ref{sec:BackgroundVar} the condition number of any matrix is bounded below by one. We therefore do not include the minimum values of $\kappa(\bfS)$ in the table. We firstly note that the maximum values of $\kappa(\bfS)$ are extremely large, with the largest value occurring for the matrix $\Rctrl$. For experiments with correlated OEC matrices, increasing $\lambda_{min}(\bfR)$ results in a decrease in the maximum value of $\kappa(\bfR)$. %This reduction is quite consistent for $\bfR_{est}$ to $\bfR_{500}$, with a much larger decrease from $\bfR_{500}$ to $\bfR_{67}$. 
 {\color{black} We note that the changes to the condition number for $\bfR_{67}$ compared to $\bfR_{500}$ are much larger than the difference in conditioning between other experiments.}
The maximum value of $\kappa(\bfR)$ for the OEC matrix $\bfR_{infl}$ is the smallest of all choices of OEC matrix. %The distribution of the number of iterations (see Figure \ref{fig:niter}) agrees with the order of $\max \kappa(\bfS)$ - 
A decrease in the maximum value of $\kappa(\bfS)$ corresponds to a distribution that has increased weight at the lower end of the spectrum for the iteration count distribution shown in Figure \ref{fig:niter}.\linebreak

We now consider the mean and the median of $\kappa(\bfS)$. Firstly we note that the values of the mean and median differ by at least one order of magnitude.  {\color{black} The distribution of $\kappa(\bfS)$ is not symmetric: it is bounded below by $1$, with very large maximum values. The mean is skewed by such outliers and we note that for a boxplot of this data (not shown) the mean does not lie within the interquartile range (IQR) of the data for all experiments other than $E_{67}$ and $E_{infl}$. Both the maximum and mean of $\kappa(\bfS)$ decrease with increasing $\lambda_{min}(\bfR)$, for correlated OEC matrices. The largest values occur for the experiment $\Ectrl$, and the smallest for the experiment $E_{infl}$. In contrast, the median is largest for the experiment $E_{infl}$, and decreasing $\lambda_{min}(\bfR)$ increases the median value of $\kappa(\bfS)$ for experiments with correlated choices of OEC matrix. Considering the deciles indicates that the spread of $\kappa(\bfS)$ across all observations reduces as more reconditioning is applied. }\linebreak %This right skew can also be seen in Figure \ref{fig:niter}. Very large values, such as the maximum values of $\kappa(\bfS)$, have more influence on the mean than values near $1$, so the mean is strongly skewed by such outliers. We note that the mean values of $\kappa(\bfS)$ are two orders of magnitude smaller than the  maximum value of $\kappa(\bfS)$ for all experiments.  If we consider the deciles of $\kappa(\bfS)$, up to the 7th decile (i.e. 70\% of observations) the value of $\kappa(\bfS)$ for each decile increases as $\lambda_{min}(\bfR)$ decreases. For the 8th to 10th decile, the value of $\kappa(\bfS)$ decreases as $\lambda_{min}(\bfR)$ decreases. This indicates that the spread of $\kappa(\bfS)$ across all observations reduces as more reconditioning is applied. %We recall that the bounds \eqref{eq:newboundgen} only provide information on the maximum and minimum possible values of $\kappa(\bfS)$.
%This change in deciles (including the median) is most evident for $E_{67}$. \linebreak %Could say something about a threshold if this isn't enough. This could suggest that the bias towards larger values of $\kappa(\bfS)$ reduces slightly as $\lambda_{min}(\bfR)$ increases. 
%In fact there is a slight reduction in the range of $\kappa(\bfS)$ values with increasing $\lambda_{min}(\bfR)$ for correlated choices of $\bfR$, as all choices of OEC matrices have a minimum values of $\kappa(\bfS)$ in the range $[1,1.00001]$. \linebreak%However, it appears that reconditioning only has a minimal impact on the range of the condition numbers. 

We have seen that introducing correlated OEC matrices improves convergence and reduces $\kappa(\bfS)$ compared to the current operational choice. Additionally, reducing the target condition number results in further improvements. This behaviour agrees with the theoretical conclusions of \citet{tabeart17a} that were summarised in Section \ref{sec:Recond}. For a linear observation operator we expect the upper bound on the condition number of the Hessian to decrease as the minimum eigenvalue of the OEC matrix, $\bfR$, increases. This is shown in \eqref{eq:newboundgen}.
This equation also shows that the ratio between background and observation variance is important for the conditioning of $\bfS$. The final column of Table \ref{table:condHess} shows that 
range of $\kappa(\bfS)$ for the experiment $E_{infl}$ is less than the range for any experiment with a correlated choice of OEC matrix. The variances for $\bfR_{infl}$ are much larger than the variances for any other OEC matrix considered in this work. 
We therefore conclude that the qualitative conclusions of \citet{tabeart17a}, as presented in Section \ref{sec:Recond}, hold in this framework, even in the case of a non-linear observation operator. 
\subsection{Effect of changing the observation error covariance matrix on 1D-Var Retrievals}\label{sec:MO1DVar}
%\textit{Need to define retrieved standard deviation: something like sqrt of diagonal of inverse of Hessian ie diagonal of analysis error covariance matrix, but I need to find the documentation page...}
%Having illustrated the impact of reconditioning on the correlation and covariance matrices, we can consider the effect of our choice of $\bfR$ on the results of the 1D-Var assimilation.  What we have referred to in this work as `ridge regression' in Algorithm $1$ is the reconditioning method that is used operationally at the Met Office. For this reason our numerical experiments and trials will focus on this case, although other methods of reconditioning were tested in a more limited fashion. The experimental design and parameter choice were described in Section \ref{sec:experimentaldesign} and the notation used in this section was introduced there. \linebreak
%I feel like this paragraph is perhaps unnecessary?

%The types of presentation that were proved of most interest can be split into two types: those which focus on the effects of changing $\bfR$ on the 1DVar retrievals, and those which consider effects on the Hessian and convergence of the minimisation problem. We will consider both of these categories in turn, firstly looking at the 1DVar retrievals. \linebreak

%1DVar retrievals:
In this section we consider how changing the OEC matrix impacts the retrieved values of physical variables. In particular we focus on temperature and specific humidity, as we obtain profiles that occur across multiple model levels rather than individual values. We note that the retrieved temperature and humidity values are not passed to the 4D-Var assimilation procedure. However, studying how these variables change for different choices of OEC matrix helps us understand the impact of changing the OEC matrix, $\bfR$, on the 1D-Var assimilation.  Additionally, as part of the 1D-Var assimilation procedure, retrieved standard deviation (RSD) values for each of the retrieval values are derived. The RSD values are calculated as the square root of the diagonal entries of the inverse of the Hessian given by \eqref{eq:Hessian}, i.e. the retrieved analysis error covariance in state variable space. For each 1D-Var assimilation we obtain a different value for RSD for each retrieved variable. We therefore consider the average RSD value for a given experiment and retrieved variable. For temperature and specific humidity this means that we obtain different RSD values for each model level. Comparing the range of differences between retrievals to the RSD values will allow us to determine whether the difference made when changing the OEC matrix, $\bfR$, is of a similar order to expected variation, or much larger (and hence results in significant differences). We will also compare RSD and differences against BSD values as shown in Figure \ref{fig:Bsd}.\linebreak %By considering the change to 1D-Var retrievals we can compare what effect our choice of observation error covariance has on the result of the minimisation of the cost function. 
%We begin by considering how changing $\bfR$ impacts temperature retrievals. % The control vector that is output contains information about $9$ different observation types. Of most interest here are the values corresponding to the temperature levels and the specific humidity because for each of these we have profiles that occur across multiple model levels rather than simply individual values. The model levels are determined by the $43$ pressure levels of RTTOV (see Section \ref{sec:MOsystem} for further details). The control vector returns $43$ levels of temperature, and $26$ levels of humidity. These $26$ levels correspond to the lowest levels of the model and approximately cover the troposphere \cite{prates16}. We begin by considering temperature. \linebreak %has 87 entries, which correspond to 9 observation types across the $43$ pressure levels of RTTOV. These entries are made up of $43$ levels of temperature, $26$ levels of $\ln(q)$ (the natural logarithm of specific humidity), the surface air temperature, the surface $\ln(q)$, surface pressure, surface skin temperature, cloud top pressure, effective cloud fraction and $12$ infrared surface emissivity principal components. Of most interest here are the values corresponding to the temperature levels and the specific humidity because for each of these we have profiles that occur across multiple model levels rather than simply individual values. \linebreak

\begin{figure*}\centering
		\includegraphics[width=0.9\textwidth]{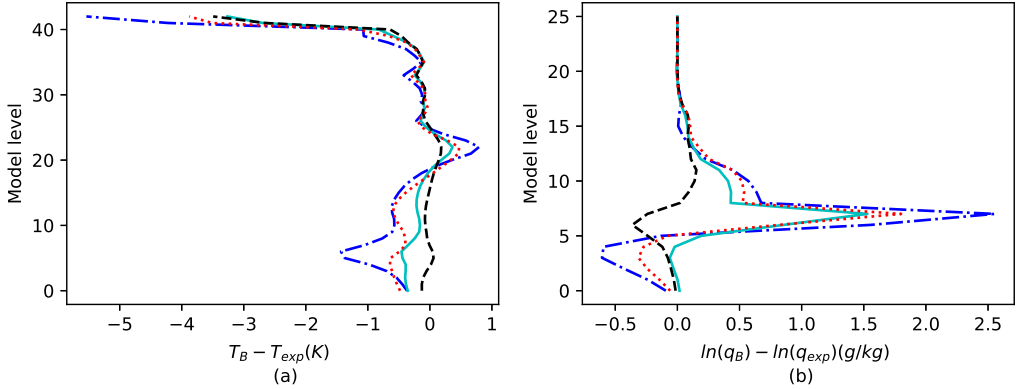}
%	\begin{subfigure}{0.48\textwidth}
%	\centering
%	\includegraphics[trim=0mm 5mm 0mm 0mm ,clip,width=\textwidth]{Fig1aTemp}
%%	\caption[Results of 1D-Var retrieval]{A temperature retrieval from 16th March 2016 0000Z. Plotted are the temperature retrievals from trials corresponding to $\bfR_{oper}$ (black solid line), $\bfR_{unpre}$ (cyan solid line), $\bfR_{500}$ (blue triangle) and $\bfR_{67}$ (red upside down triangle). The observation is for a latitude of $-20.0342$ and longitude of $4.5961$.}
%	\label{fig:1DVarRetrieval}
%	\end{subfigure}
%	\begin{subfigure}{0.48\textwidth}
%		\centering
%		\includegraphics[trim=0mm 5mm 0mm 0mm, clip,width=\linewidth]{Fig1bHum}
%		\label{fig:Qretrieval}	
%	\end{subfigure}	

		\caption{Background minus retrieved profiles from observation at (-33.16N,-32.70E) for 16th June 2016 0000Z for (a) temperature (b) ln(specific humidity). Differences are shown for $\Ectrl$ (dot-dashed line), $E_{est}$ (dotted line), $E_{67}$ (solid line) and $E_{infl}$ (dashed line). \label{fig:1DVarRetrieval}} %\todo[inline]{Need to redo these plots with same linestyles. Also probably expand on caption}} %Her
\end{figure*}
Figure \ref{fig:1DVarRetrieval} shows {\color{black}background profiles minus retrieved profiles} for temperature and humidity for observations at the location (-33.16N,-32.70E). Retrievals are shown at pressure levels in the atmosphere. These model levels are determined by the $43$ evenly distributed pressure levels in the radiative transfer retrieval algorithm.  Specific humidity is only calculated for the lowest $26$ model levels. We note that this is the configuration that was used at the time of the experiments (July 2016). {\color{black}Differences from the background are larger for specific humidity profiles than for temperature. However qualitative behaviour is similar for both variables. In both cases $E_{diag}$ is the most different from the background, implying that the use of correlated observation errors increases the weighted importance of the background. {\color{black} Increasing the amount of reconditioning used decreases the norm of the difference between the retrieved profile and the background for all correlated OEC matrices. Hence, applying a larger amount of reconditioning results in a retrieved profile that is closer to the background.}
%We also see that reconditioning results in a retrieved profile that is closer to the background. 
Finally, the retrieval corresponding to $E_{infl}$ is closest to the background for both variables. For this case, standard deviations have been inflated, meaning that we expect the retrieved profile to fit closer to the background. This is particularly evident for specific humidity where there is a large difference between background and retrieved values for model level 7 for $\Ectrl, E_{est} $ and $E_{67}$. This occurs due to large differences between background and retrieved brightness temperature for channels 128-137, which have water vapour mixing ratio Jacobians that peak at pressure level 7 \citep{stewart09}. We recall that these channels are sensitive to water vapour, and have the strongest positive correlations in $\bfR_{est}$.  This explains why specific humidity is particularly affected by changes to the OEC matrix for this model level, although we note that noticeable changes also occur for temperature for this model level. }\linebreak

\begin{figure*}
%	\begin{subfigure}{0.48\textwidth}
%		\centering
%		\includegraphics[width=\linewidth]{rawdiffDesroziersMarch}
%		\caption{$\bfR_{oper}-\bfR_{unpre}$}
%		\label{fig:rawdiffDesroziers}
%	\end{subfigure}	
%	\begin{subfigure}{0.48\textwidth}
%		\centering
%		\includegraphics[width=\linewidth]{rawdiff67March}
%		\caption{$\bfR_{oper} -\bfR_{67}$}
%		\label{fig:rawdiff67}
%	\end{subfigure}		
\centering
		\includegraphics[trim=5mm 10mm 5mm 5mm, clip,width=\textwidth]{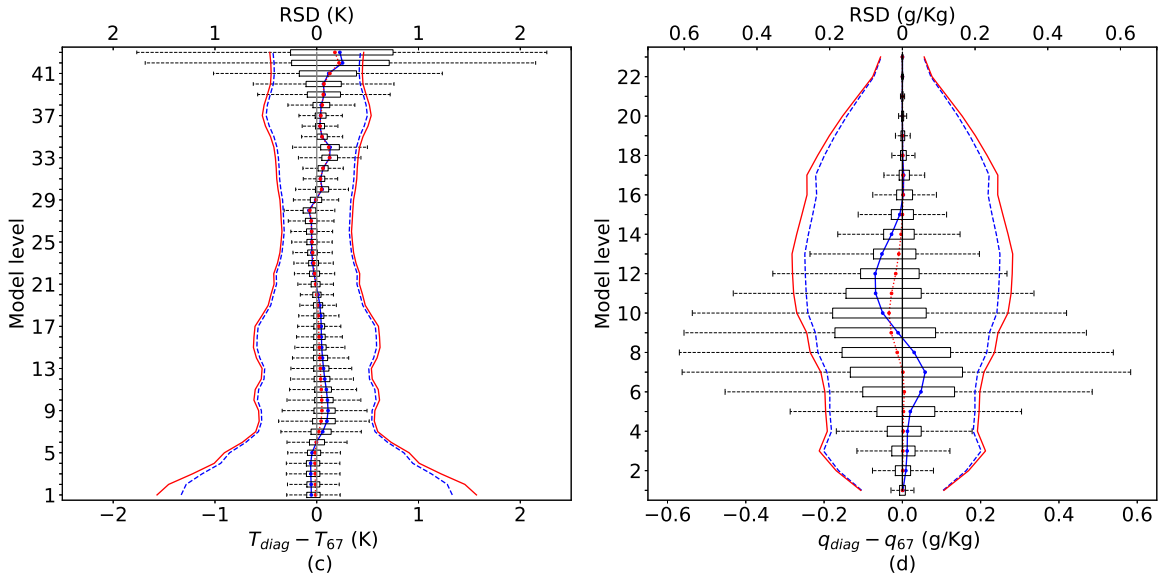}
	\caption[Differences in retrievals]{Differences in retrievals  between $\Ectrl$ and $E_{67}$ for trial on 16th June 0000Z for (a) temperature and (b) ln(specific humidity) for 97330 observations. Dashed lines and solid lines give the mean RSD values for $\Ectrl$ and $E_{67}$ respectively. Dashed lines with dots denote the median and solid lines with dots denote the mean for each pressure level. The solid box contains the middle $50\%$ of the data, and the whiskers (dashed horizontal lines) extend to the quartiles plus/minus $1.5$ times the interquartile range (IQR) - the  difference between the third and first quartiles. Outliers, which lie outside the range of the whiskers, are not shown. } % \todo[inline]{Only want one out of this and normalised figure - pick! Also formatting is grim, will probs need to redo. Could we get away with one plot and saything others are similar? Then combine temp on left and specific humidity on right?}} %}
	% Figure produced using Comp_June_Dataset.py on harddrive
	\label{fig:rawdiffall}
\end{figure*}

 %figures for the other correlated choices of $\bfR$ were very similar, and hence are not shown here.  %, along with the mean of the two corresponding standard deviations. 
%For each of the $43$ model levels a box plot of these differences was then plotted. 
 %The influence of the outliers (not shown here) can be seen by the large differences in the mean  and the median. %The 1D-Var routine also retrieves a standard deviation value (RSD) for each of the physical retrieved variables.
 % The average value of these is plotted as a red solid line for the RSD corresponding to $E_{67} $ and blue dashed line for the RSD corresponding to $\Ectrl$.
We now consider the differences between retrieved values for $\Ectrl$ and $E_{67}$ for all $97330$ observations that were accepted by the 1D-Var routine for all choices of OEC matrix. Figures \ref{fig:rawdiffall}a and b are box plots showing the distribution of these differences across each model level for temperature and ln(specific humidity profiles) respectively.% for $\Ectrl$ and $E_{67}$ for each of the $97330$ observations that were accepted by the 1D-Var routine for all choices of OEC matrix. %For each of the $98475$ observations that are accepted by the 1D-Var routine for all choices of $\bfR$,  the difference was calculated between the retrieved temperatures for two cases: $\bfR=\bfR_{oper}$ (uncorrelated), and a choice of correlated $\bfR$.
  The qualitative behaviour for other experiments was very similar and is not shown here. 
 Figures \ref{fig:rawdiffall}a shows that for most model levels the whiskers are contained within the average RSD values, and for model levels $1-41$ the central $50\%$ of differences lie within the averaged RSD.   This indicates that changing from an uncorrelated to correlated choice of OEC matrix has a generally small impact on temperatures for the majority of model levels compared to RSD.\linebreak%, but a large impact for model levels $42$ and $43$.  
%Indeed, for the majority of channels, the change to retrieved temperatures is small compared to the range of temperatures observed throughout the atmosphere. %Phrase this better.

%We now consider the mean and median of the differences between retrieved temperature values, shown in Figure \ref{fig:rawdiffall}a. 
%Although for many model levels the mean and median are non-zero, the two values are close for most cases.  %The largest differences in mean and median are seen for model levels $8$ to $14$.
 %This suggests that the differences between retrievals may have a symmetric distribution, with non-zero mean. %This could be due to bias or systematic error introduced by  correlated choices of OEC matrices, in which case it may be possible to remove or eliminate this error.  
Mean RSD values for $E_{67}$  and $\Ectrl$ are also very similar, with %, with the largest differences occurring for model levels 1 and 2.  
larger mean RSD values for $E_{67}$ than $\Ectrl$ for all model levels. This is observed for all correlated choices of OEC matrix; the mean RSD is increased for all model levels compared to $\Ectrl$. This suggests that using a correlated choice of OEC matrix increases the mean RSD for temperature i.e. by introducing correlations we have less confidence in the retrieved values, or 1D-Var analysis. This increase to standard deviations is expected from theoretical and idealised studies \citep{stewart08,rainwater15,fowler17}. We also note that by including correlations we put less weight on the individual channels but allow more freedom to fit  multivariate information arising from the combination of channels. {Comparing the RSD values to the BSD values given by Figure \ref{fig:Bsd} we find that across all three choices of $\bfB$, the standard deviation values are similar to RSD values for most model levels. For model levels where the BSDs are smaller than both $E_{diag}$ and experimental RSD, differences are small in comparison to all standard deviation values.  }\linebreak

Figure \ref{fig:rawdiffall}b shows the differences between retrieved values of specific humidity for $\Ectrl$ and $E_{67}$ for $23$ model levels. As was the case for temperature, the mean RSD values for all other choices of experiment are larger than those for $\Ectrl$. We note that differences for model levels $1$ and $18-23$ are very small compared to RSD. However, for model levels $5- 12$, the whiskers lie outside the values for mean RSD. This means there is a large proportion of model levels where changing the OEC matrix has a larger impact on retrieved specific humidity values than we would expect due to instrument noise and other quantified types of uncertainty. We also note that for these model levels we have non-zero and non-equal means and medians. This suggests that the distribution of differences is not symmetric. Again, BSD values are larger than RSD values for the majority of model levels for specific humidity. However, whiskers still extend past the BSD values for levels 5 - 10 for all choices of $\bfB$.  \linebreak

The effect of changing the OEC matrix, $\bfR$, seems to affect a larger proportion of the retrieved specific humidity values than temperature values. This coincides with the findings of \citet{bormann16,weston14}. They found large changes to humidity fields with the introduction of correlated OEC matrices in 4D-Var assimilation procedures, which resulted in improved NWP skill scores. %of high levels of spatial and interchannel correlation for channels which are highly sensitive to water vapour \citep{bormann10,stewart08b,weston11,weston14} 
\section{Impact on variables that influence the 4D-Var routine}\label{sec:Results4DVar}
In Section \ref{sec:Results1DVar} we showed that the choice of OEC matrix, $\bfR$, does make a difference to the 1D-Var routine in terms of convergence, and the individual retrieval values. We now consider variables that directly impact the main 4D-Var procedure that is used to initialise forecasts. Changes to the OEC matrix in the 1D-Var routine affect 4D-Var in two main ways: firstly by altering the observations that are accepted by the quality control procedure, and secondly via retrieved values of variables that are not analysed in the 4D-Var state vector. We will consider these two aspects in turn.

\subsection{Changes to the quality control procedure}\label{sec:ExptvsCtrl}

%We are interested in the effect of changing the observation error covariance matrix in the 1D-Var assimilation procedure at the Met Office. \todo{Should the following sentence go in Section 3?} This procedure is carried out as part of a quality control system, as observations that converge too slowly in a 1D-Var framework are thought to be likely to cause poor convergence in a 4D-Var framework. Additionally the 1D-Var assimilation fixed skin temperature (ST), cloud top pressure (CTP) and cloud fraction (CF) variables which are not changed by the 4D-Var assimilation. In this section we focus on how changing the observation error covariance matrix in 1D-Var changes number of accepted observations, and the three fixed variables. We recall that we consider $7$ choices of observation error covariance matrix: $ \bfR_{oper}$, used in the current 1D-Var assimilation framework at the Met Office which is diagonal and uninflated, $\bfR_{raw}$ which is the symmetrised version of the Desroziers diagnostic with no reconditioning applied, $\bfR_{1500}$, $\bfR_{1000}$, $\bfR_{500}$ and $\bfR_{67}$ reconditioned matrices where the subscript denotes the condition number of the $137\times 137$ correlated submatrix, and $\bfR_{old}$ the previous choice of $\bfR$ for 1D-Var which is diagonal and inflated.
%\subsubsection{Number of observations that are accepted}\label{sec:NoObsAccepted}
 In Section \ref{sec:MOHessian} we showed that increasing $\lambda_{min}(\bfR)$ increases the speed of convergence of the 1D-Var routine. We now investigate whether changing the OEC matrix, $\bfR$, alters the number of observations that pass the quality control step that was described in Section \ref{sec:MOsystem}. We also consider how the number of observations accepted by experiment (respectively $E_{diag}$) and rejected by $E_{diag}$ (respectively experiment) changes for different choices of OEC matrix.  This information is presented in Table \ref{table:acceptedobs}.\linebreak
 
 We begin by considering in more detail why changing the OEC matrix would result in changes to the number of observations that pass quality control. 
 Observations are rejected if the minimisation of the 1D-Var procedure requires more than 10 iterations to converge. In Section \ref{sec:MOHessian} we found that introducing correlated observation error reduces the number of iterations required for convergence, and that decreasing the target condition number increases convergence speed further. This suggests that introducing correlated OEC matrices and using reconditioning will result in a larger number of observations that converge fast enough to pass this aspect of quality control. We therefore expect the use of reconditioning methods to result in a larger number of accepted observations. \linebreak

%\cite{tabeart17a} demonstrated that for a linear data assimilation problem, increasing $\lambda_{min}(\bfR)$ reduces the number of iterations required for convergence. In Section \ref{sec:MOHessian} we showed that increasing $\lambda_{min}(\bfR)$ increases the speed of convergence of the 1D-Var routine. We now investigate whether observations that take more than the maximum $10$ iterations for the control require fewer iterations for other choices of $\bfR$. We also consider how the raw numbers of accepted observations changes for each choice of $\bfR$. This information is shown in Table \ref{table:acceptedobs}.

\begin{table*}
	\centering
	\begin{tabular}{ lc c  c c c c  }
		\hline
		&\multicolumn{6}{c}{ $E_{exp}$}\\
		\hline
		Experiment & $E_{est}$ & $E_{1500}$ & $E_{1000}$ & $E_{500}$ & $E_{67}$ &$ E_{infl}$ \\
		\hline
		No. of accepted obs (T) & $100655$&$ 100795$& $101002$&$ 101341$& $102333$& $102859$ \\
		No of obs accepted by both $\Ectrl$ and $E_{exp}$ & $99039$&$ 99175$&$99352$& $99656$& $100382$& $100679$ \\
		Accepted by $E_{exp}$, rejected by $\Ectrl$ & $1616$ &$1620$& $1650$& $1685$& $1951$&$2180$ \\
	%	\% of T accepted by $E_{exp}$, rejected by $\Ectrl$ &1.61\%&1.61\%&1.63 \%&1.66\%&1.91\%&2.12\%\\
		Accepted by $\Ectrl$, rejected by $E_{exp}$ &$1647$&$1511$&$1334$&$1030$&$304$&$7$\\
	%	\% of T accepted by $\Ectrl$, rejected by $E_{exp}$ &$1.64\%$&$1.5\%$&$1.32\%$&$1.02\%$&$0.3\%$&$0.007\%$\\
		\hline
	\end{tabular}
	\caption{Number of observations accepted by the 1D-Var quality control for each experiment ($E_{exp}$) compared to $\Ectrl$. For $\Ectrl$ the total number of accepted observations is 100686. Here T refers to the total number of distinct observations (defined in Section \ref{sec:MOsystem}) accepted by $E_{exp}$ for each experiment. The number of observations accepted by all experiments is 97330.}
	\label{table:acceptedobs}
	
\end{table*}
%We might expect an increase in the number of accepted observation for choices of $\bfR$ with faster rates of convergence than $\Rctrl$. 
The first row of Table \ref{table:acceptedobs} shows that the number of accepted observations increases as $\lambda_{min}(\bfR)$  (see Table \ref{tab:mineig}) increases  and the largest number of accepted observations occurs for experiment $E_{infl}$. This coincides with what we would expect due to alterations in the quality control procedure. However, we note that the number of accepted observations is slightly larger for $\Ectrl$ than $E_{est}$ even though convergence for $E_{est}$ was faster than for $\Ectrl$ across the set of common observations.  
%One possible reason for this is a larger number of observations failing the cloud detection tests for $E_{est}$ than $\Ectrl$. {\color{black} }. For most channels, the standard deviation corresponding to $\bfR_{diag}$ is larger than for $\bfR_{est}$. It is likely that observation-minus-background information will therefore be given a larger weight in the cloud detection test for $E_{est}$, and therefore that the cloud threshold will be exceeded and the observation rejected. \halfblankline%The reason for this behaviour is not understood.  We note that other cloud detection tests consider the difference between observations from IASI and observations from other independent instruments, as well as background information in observation space. IASI observations that differ by large amounts from these other sources of information are rejected. It is therefore possible that a larger number of observations are rejected due to large differences for $E_{est}$ compared to $E_{ctrl}$. This will be discussed further in Section \ref{sec:FixedVariablesOverall}.\linebreak%Do we have any hypothese for this? For 5.2 we see largest difference and general difference in behaviour for Rold and Rctrl - produce unphysical values?
{\color{black} The second row of Table \ref{table:acceptedobs} shows that most observations are accepted by both $E_{exp}$ and $E_{diag}$. We see that the number of accepted observations increases with $\lambda_{min}(\bfR)$ for correlated choices of $\bfR$. The largest number of observations is accepted by  $E_{infl}$. The third and fourth rows of  Table \ref{table:acceptedobs} shows the number of observations that are accepted by $E_{exp}$ (respectively $E_{diag}$) and rejected by $E_{diag}$ (respectively $E_{exp}$).
However, this number is smaller than 2.2\% of the total number of observations all choices of $E_{exp}$.} For what follows we shall consider the large majority of observations that are accepted by both  $E_{diag}$ and $E_{exp}$. Although observations that are accepted by only one of $E_{diag}$ and $E_{exp}$ are of interest, the fact that there are very few observations in either of these sets makes it hard to study their properties statistically.
\subsection{Changes to retrieved values for variables that are not included in the 4D-Var control vector}\label{sec:FixedVariablesOverall}
%Interesting facts about boxplots
%The box plot allows quick graphical examination of one or more data sets. Box plots may seem more primitive than a histogram or kernel density estimate but they do have some advantages. They take up less space and are therefore particularly useful for comparing distributions between several groups or sets of data (see Figure 1 for an example). Choice of number and width of bins techniques can heavily influence the appearance of a histogram, and choice of bandwidth can heavily influence the appearance of a kernel density estimate.
In this Section we consider how altering the OEC matrix used in the 1D-Var routine alters the retrieved values of variables that are not included in the 4D-Var control vector. {\color{black} For all three variables, Figure \ref{fig:FixedvariablesRaw} shows that the majority of  retrievals are changed by a small amount for each choice of experiment. The largest differences occur between $E_{diag}$ and $E_{exp}$ for ST, CF and CTP, where the IQR and whiskers are much larger than for any correlated choice of OEC matrix. For correlated OEC matrices, we see a reduction in IQR and whisker length as $\lambda_{min}(\bfR)$ increases. This indicates that as we increase the amount of reconditioning that is applied, the differences between $E_{diag}$ and $E_{exp}$ reduce. However, there are some differences between the variables. \linebreak %We note that we investigate only the 97330 observations that are accepted by all experiments and $E_{diag}$ and study their distributions and average behaviour.   
%We study the difference between retrieved values for skin temperature (ST), cloud top pressure (CTP) and cloud fraction (CF) for observations that are accepted by both $E_{diag}$ and experiment. \linebreak%Figure \ref{fig:FixedvariablesRaw} shows the difference between retrievals for these three variables for $\Ectrl$ and the other experiment $E_{exp}$.\linebreak 

\begin{figure*}
	\centering
		\includegraphics[ width=\textwidth]{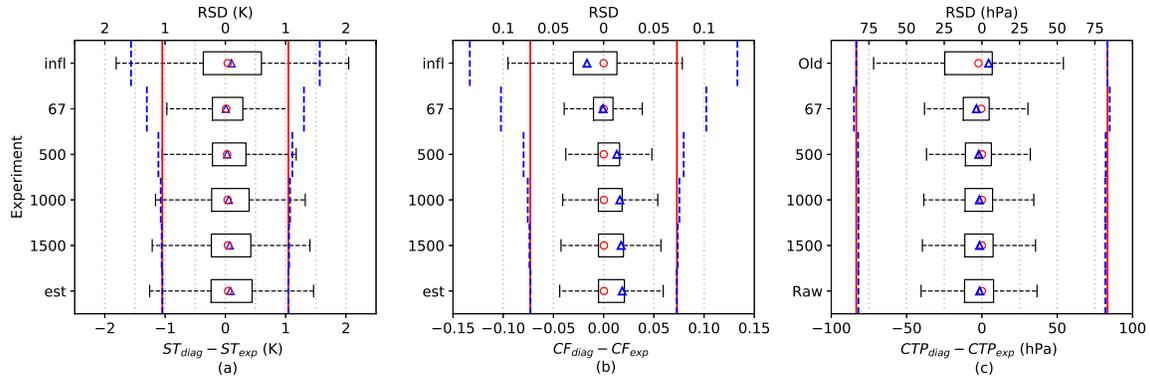}

	\caption{Box plot showing differences between retrieved variables for $\Ectrl-E_{exp}$ for (a) ST (skin temperature) (b) CF (cloud fraction) and (c) CTP (cloud top pressure). The  circle shows the median, the triangle depicts the mean, the solid  box contains the central $50\%$ of data (the interquartile range), and the dashed horizontal lines show the whiskers which extend to the quartile $\pm1.5\times IQR$. Vertical dashed lines show the mean retrieved standard deviation (RSD) values for the experiment, and the solid vertical lines shows the mean RSD values for $E_{diag}$. Outliers (not shown) lie in the range (a) $\pm 33.52K$, (b) $\pm 1$ and (c) $\pm 913.25hPa$. {\color{black} The number of outliers and extreme outliers for these experiments is presented in Tables \ref{table:outlierobs} and \ref{table:extremeobs}}.}
	\label{fig:FixedvariablesRaw}
\end{figure*}

Firstly, for ST all choices of OEC matrix yield whiskers that are equal to or exceed the RSD values corresponding to $E_{diag}$ (solid line), and all except $E_{67}$ exceed the RSD values for the corresponding experiment (dashed line). In contrast, the whiskers for correlated choices of OEC matrix are well within both RSD values for CF and CTP, as well as the BSD values given in Table \ref{tab:Bsd}.  This shows that compared to expected observation variability, differences between CF and CTP retrievals are small for correlated choices of OEC matrix. However, we recall that BSD values for cloud variables were artificially inflated \citep{Pavelin08}.\linebreak

For ST and CTP the values of the mean and median are close for all correlated choices of $E_{exp}$, and the box and whiskers are fairly symmetric about $0$. In contrast, for CF, differences between the mean and median occur, and the box extends further into the positive axis. Cloud errors are expected to vary greatly with the cloud state, meaning that it is difficult to interpret gross statistics \citep{Eyre89}. We include them here for completeness.   }\linebreak

\begin{table*}
	\centering
	\begin{tabular}{ l   c c  c c c c  }
		\hline
		&  $E_{est}$ & $E_{1500}$ & $E_{1000}$ & $E_{500}$ & $E_{67}$ &$ E_{infl}$ \\
		\hline
		\% outliers (ST) &15.1&15.3&15.6&16.3&17.6 &15.9\\
		\% of outliers (CF) &23.9&24.02& 24.2& 24.6& 25.3& 21.4 \\
		%	Maximum difference (CF) &$-$ &1.61\%&1.61\%&1.63 \%&1.66\%&1.91\%&2.12\%\\
		%	Minimum difference (CF) &$-$& $99039$&$ 99175$&$99352$& $99656$& $100382$& $100679$ \\				
		\% outliers (CTP) &22.8&22.8&23.0&22.9&21.4&18.8\\
	%No. of positive outliers (CTP) & 10584 & 10591& 10642 & 10589 & 9839& 9584 \\
	%No. of negative outliers (CTP)&11562 & 11644& 11702& 11688& 10986& 8743 \\

		\hline
		Maximum difference (ST (K)) &21.67& 21.12& 21.14& 22.38& 21.03& 26.83\\
		Minimum difference (ST (K)) & -33.52& -33.01& -32.14& -29.76& -23.82& -20.88\\
		%	Maximum difference (CTP) &$-$ &1.61\%&1.61\%&1.63 \%&1.66\%&1.91\%&2.12\%\\
		%	Minimum difference (CTP) &$-$& $99039$&$ 99175$&$99352$& $99656$& $100382$& $100679$ \\
		
		\hline
	\end{tabular}
	\caption{Percentage of outliers for cloud fraction, cloud top pressure and skin temperature. Outliers are differences which fall outside the whiskers shown in Figure \ref{fig:FixedvariablesRaw}. Maximum and minimum differences are shown for skin temperature only; maximum differences for cloud fraction and cloud top pressure are $\pm1$ and $\pm913.25hPa$ respectively, for all choices of $\bfR$.}
	\label{table:outlierobs}
	
\end{table*}

%\begin{table*}
%	\centering
%	\begin{tabular}{ l   c c  c c c c  }
%		\hline
%		&  $E_{est}$ & $E_{1500}$ & $E_{1000}$ & $E_{500}$ & $E_{67}$ &$ E_{infl}$ \\
%		\hline
%				No. of extreme outliers ($|ST| >5K$) & 1556& 1501& 1435& 1336& 1393& 3552 \\
%		No. of extreme outliers ($|CF| >0.25$) &4748& 4543& 4327& 3827& 3112& 7331 \\
%		No. of extreme outliers ($|CTP|> 225hPa$) & 3244& 3245& 3230& 3167& 2674& 4261 \\
%
%		%	Maximum difference (CTP) &$-$ &1.61\%&1.61\%&1.63 \%&1.66\%&1.91\%&2.12\%\\
%		%	Minimum difference (CTP) &$-$& $99039$&$ 99175$&$99352$& $99656$& $100382$& $100679$ \\
%		
%		\hline
%	\end{tabular}
%	\caption{Number of extreme outliers for cloud fraction, cloud top pressure and skin temperature for each experiment. Extreme outliers are defined as observations with absolute differences greater than $0.25$ for CF, $225hPa$ for CTP and $5K$ for ST. This corresponds to absolute differences greater than approximately $25\%$ of the maximum differences presented in Table \ref{table:outlierobs}. }
%	\label{table:extremeobs}
%	
%\end{table*}
\begin{table*}
	\centering
	\begin{tabular}{ l   c c  c c c c  }
		\hline
		&  $E_{est}$ & $E_{1500}$ & $E_{1000}$ & $E_{500}$ & $E_{67}$ &$ E_{infl}$ \\
		\hline
		\% extreme outliers ($|ST|>5K$) & 1.6&1.5&1.5&1.4&1.4&3.6\\
	\% extreme outliers ($|CF| >0.25$)& 4.9&4.7&4.4&3.9&3.2&7.5 \\
%	No. of extreme outliers ($|CF| >0.25$) &4748& 4543& 4327& 3827& 3112& 7331 \\
\% extreme outliers ($|CTP| >225hPa$) &3.3&3.3&3.3&3.3&2.7&4.4\\
%      				No. of extreme outliers ($|CTP|> 225hPa$) & 3244& 3245& 3230& 3167& 2674& 4261 \\

		%	Maximum difference (CTP) &$-$ &1.61\%&1.61\%&1.63 \%&1.66\%&1.91\%&2.12\%\\
		%	Minimum difference (CTP) &$-$& $99039$&$ 99175$&$99352$& $99656$& $100382$& $100679$ \\
		
		\hline
	\end{tabular}
	\caption{Number of large outliers for cloud fraction, cloud top pressure and skin temperature for each experiment. Large outliers are defined as observations with absolute differences greater than $0.25$ for CF, $225hPa$ for CTP and $5K$ for ST. This corresponds to absolute differences greater than approximately $25\%$ of the maximum differences presented in Table \ref{table:outlierobs}. }
	\label{table:extremeobs}
	
\end{table*}

For all variables, the majority of retrievals change by a small amount, relative to RSD, when comparing the experiment to $E_{diag}$. %This means that the majority of observations experience small effects when changing the OEC matrix.   The box and whisker plots provide us with 'typical' values.
However, {Table \ref{table:outlierobs} shows that \color{black}over 15\% of observations are classed as outliers for all three variables. These outliers are defined as observations with retrieval differences that are not between $Q_1-1.5IQR$ and $Q3 + 1.5IQR$, where $Q_1$ and $Q_3$ denote the first and third quartiles of the data respectively, and are not shown in Figure \ref{fig:FixedvariablesRaw}. Not all of these outliers represent large differences between retrieved values. Instead we consider `large' outliers, which we define in this setting as differences larger than $25\%$ of the maximum differences for each variable.
For cloud variables the maximum difference is defined by the possible range of values: $\pm1$ and $\pm913.25hPa$ for CF and CTP respectively. For ST we use the maximum difference between retrievals from the data set. These values are given in Table \ref{table:outlierobs}. \linebreak %We note that the maximum absolute difference of 33.52K is unphysical, and that only a single observation takes this value. \linebreak

Table \ref{table:extremeobs} shows the percentage of large outliers for each variable, which is  %, where we can see that the number of extreme differences is much smaller than the number of outliers. 
much smaller than the total number of outliers for all variables and experiments. For all variables, the number of large outliers decreases with $\lambda_{min}(\bfR)$ for correlated experiments. The experiment $E_{infl}$ has a much greater number of large outliers than any experiment with a correlated choice of OEC matrix, agreeing with earlier findings that the qualitative and quantitative differences between $E_{infl}$ and $\Ectrl$ are much larger than for any other experiment. } \linebreak% As the 1D-Var system is tuned to provide good results for $E_{ctrl}$, it is possible that these large differences between meteorological variables for accepted values are even larger for rejected observations.

{\color{black} As background information has almost no weight for cloud variables, due to inflated BSD values, changing the OEC matrix could result in much larger differences between retrieved values for CF and CTP than for other variables. However, this is not the case for ST, where the maximum differences given in Table \ref{table:outlierobs} are extremely large compared to RSD and BSD values. The number of observations with extremely large retrievals is small: for correlated experiments fewer than 10 observations yield absolute differences larger than 20K. These observations can be considered as failures of the 1D-Var algorithm and should be removed by the quality control procedure. This emphasises that when altering the OEC matrix, the quality control procedure needs to be altered as well.} \linebreak

  Previous studies by \citet{stewart14,weston14,bormann16,campbell16} have shown that the largest impacts of applying the DBCP diagnostic to IASI occur for humidity sounding channels, which will affect clouds and retrieved values associated with clouds.  Skin temperature is also sensitive to cloud; although in partly overcast conditions it is possible to retrieve estimates of skin temperature, errors in the modelling of cloud effects are likely to dominate the surface signal \citep{stewart14,Pavelin14}. In terms of impact, under cloudy conditions the 4D-Var assimilation procedure is less sensitive to skin temperature \citep{Pavelin14}, so it is possible that these large changes to retrievals will not result in large  impacts when passed to 4D-Var. However, further work is needed to understand the origin and consequences of these extreme differences fully.

\section{Conclusions}\label{sec:Conclusion}
It is widely known that many observing systems in numerical weather prediction (NWP) have errors that are correlated \citep{janjic17} for reasons including scale mismatch between observation and model resolution, approximations in the observation operator or correlations introduced by preprocessing. However, diagnosed error covariance matrices have been found to be extremely ill-conditioned, and cause convergence problems when used in existing NWP computer systems \citep{campbell16,weston14}. \citet{tabeart17a} established that increasing the minimum eigenvalue of the OEC matrix improves bounds on the conditioning of the associated linear variational data assimilation problem. This provided insights into possible reconditioning methods which could permit the inclusion of correlation information while ensuring computational efficiency \citep{tabeart17c}.\linebreak %Reconditioning methods \cite{tabeart17c} provide insight into methods which could allow us to include correlation information in a computationally efficient manner.

In this paper we have investigated the impact of changing the OEC matrix for the IASI instrument in the Met Office 1D-Var system, an operational non-linear assimilation system. In particular we have considered how reconditioning methods could permit the implementation of correlated observation error matrices. The 1D-Var system is used for quality control purposes and to retrieve values of variables that are not included in the 4D-Var state vector. As each observation is assimilated individually, it is more straightforward to understand and isolate the effects of using different choices of OEC matrix on retrieved variables and convergence compared to the more complicated 4D-Var procedure.\linebreak %This means that it is simpler to isolate the impact of the changes% want to say something like it’s a simpler to understand the effect of implementing correlated ob error on retrievals/easier to attribute or isolate impacts
%an operational non-linear assimilation system.

We found that:
\begin{itemize}
\item The current operational choice of observation error covariance (OEC) matrix for IASI results in the slowest convergence of the 1D-Var routine of all OEC matrices considered. Increasing the amount of reconditioning applied to correlated OEC matrices improves convergence of the 1D-Var routine, in accordance with the qualitative theoretical conclusions of \citet{tabeart17a,tabeart17c}.
	
%\item Changing the observation error covariance matrix resulted in a small effect on retrieved temperature values, but larger impact on retrieved humidity values.
%\item Increasing the minimum eigenvalue of the observation error covariance matrix reduces number of iterations to required for  convergence of the 1D-Var routine, and reduces condition number of the Hessian of the 1D-Var cost function.
%\item Reducing observation error variance worsens convergence of the 1D-Var routine significantly. In particular the operational choice of uncorrelated error covariance was shown to have a worse performance using than the most ill-conditioned correlated error covariance matrix. %This was shown to be an effect of smaller variance in the observation error.
	\item Most experimental choices of correlated OEC matrix resulted in a larger number of IASI observations that were accepted by the 1D-Var routine than the current diagonal operational choice. Increasing the amount of reconditioning applied to correlated OEC matrices increases the number of IASI observations that converge in fewer than 10 iterations, and hence pass the quality control component of 1D-Var.
	\item   Retrieval differences for skin temperature, cloud fraction and cloud top pressure are smaller than retrieved standard deviation values for over 75\% of IASI observations for all choices of correlated OEC matrix. Up to 5\% of retrievals have large differences relative to the retrieved standard deviation.  %DO we want to say something about why this might be?
	\item As the minimum eigenvalue of the OEC matrix is increased, the difference between $E_{diag}$ (using the current operational diagonal OEC matrix) and experimental retrieved values reduces.  %Further investigation is being carried out into the effects of using correlated R in the 1D-Var routine on the outputs of the 4D-Var routine. This will be the subject of a future paper. (Also a limitation)
\end{itemize}
{\color{black} We also find that for most variables studied RSD values are of a similar size to BSD values. %, with BSD values being larger than RSD values for all three choices of background error covariance matrix for variables not included in the 4D-Var state vector. 
We note that the BSD values for cloud variables are artificially inflated, and are hence an order of magnitude larger than the corresponding RSD values. This indicates that observation information has as large or a larger weight in the 1D-Var objective function than background profiles. \linebreak}

The qualitative conclusions from this work agree with the theoretical results of \citet{tabeart17a}, which prove that for a linear observation operator, increasing the minimum eigenvalue of the OEC matrix is important in terms of convergence of a variational data assimilation routine.\linebreak
 
  We emphasise that these convergence results contradict the common assumption that the use of correlated OEC matrices in a variational data assimilation scheme will cause convergence problems. In fact, one key benefit of using correlated OEC matrices in a 1D-Var framework is the increase in convergence speed, particularly when combined with reconditioning methods. 
  At the Met Office the 1D-Var routine is run every 6 hours for the global model so reducing the cost of the routine would save significant computational effort. Additionally, the faster convergence that is achieved by correlated choices of OEC could permit stricter convergence criteria, e.g. reducing the maximum number of iterations from $10$ to $8$, which would also result in computational savings. However, care needs to be taken to consider how this will interact with other aspects of the quality control procedure and ensure that `good' observations are not rejected.  \linebreak
  
  Changes to OEC matrices also alter the quality control aspect of the 1D-Var procedure, so care needs to be taken to ensure that  these changes system are well understood. In particular, reducing the number of iterations required for convergence of the 1D-Var routine means that a larger number of observations were accepted by our tests and passed to the 4D-Var routine. For observations that were accepted by all experiments, we considered changes to retrieved estimates for skin temperature, cloud top pressure and cloud fraction. 
  Although changes to the retrieved values with different OEC matrices were small for the majority of observations, for a small percentage of observations, the differences between retrieved values were very large.
  As ST, CTP and CF are not estimated as part of the 4D-Var procedure, such large changes may have significant effects on the analysis for 4D-Var.  {\color{black} The most extreme of these differences (particularly for ST) are unrealistic and can be viewed as 1D-Var failures. This highlights that changes to the 1D-Var system, such as with the introduction of correlated OEC matrices, must be made in conjunction with tuning of the quality control procedures.} In general, improvements to convergence need to be be balanced with impacts on other aspects of the assimilation system, such as changes to quality control, analysis fit and forecast skill.
  
\section*{Acknowledgements}

This work is funded in part by the EPSRC Centre for Doctoral Training in Mathematics of
Planet Earth, the NERC Flooding from Intense Rainfall programme (NE/K008900/1), the EPSRC
DARE project (EP/P002331/1) and the NERC National Centre for Earth Observation.
\bibliographystyle{plainnat}
%\addbibresource{}
%\bibliography{/home/jemima/Dropbox/UoR/PhD/Papers/MOpaper/bibPaper1DVar}
\bibliography{bibPaper1DVar}

\begin{thebibliography}{44}
\providecommand{\natexlab}[1]{#1}
\providecommand{\url}[1]{\texttt{#1}}
\expandafter\ifx\csname urlstyle\endcsname\relax
  \providecommand{\doi}[1]{doi: #1}\else
  \providecommand{\doi}{doi: \begingroup \urlstyle{rm}\Url}\fi

\bibitem[Bathmann(2018)]{bathmann18}
K.~Bathmann.
\newblock {Justification for estimating observation-error covariances with the
  Desroziers diagnostic}.
\newblock \emph{Quarterly Journal of the Royal Meteorological Society},
  144\penalty0 (715):\penalty0 1965--1974, 2018.
\newblock \doi{10.1002/qj.3395}.
\newblock URL
  \url{https://rmets.onlinelibrary.wiley.com/doi/abs/10.1002/qj.3395}.

\bibitem[Bennitt et~al.(2017)Bennitt, Johnson, Weston, Jones, and
  Pottiaux]{QJ:QJ3097}
G.~V. Bennitt, H.~R. Johnson, P.~P. Weston, J.~Jones, and E.~Pottiaux.
\newblock {An assessment of ground-based GNSS Zenith Total Delay observation
  errors and their correlations using the Met Office UKV model}.
\newblock \emph{Quarterly Journal of the Royal Meteorological Society}, 2017.
\newblock ISSN 1477-870X.
\newblock \doi{10.1002/qj.3097}.
\newblock URL \url{http://dx.doi.org/10.1002/qj.3097}.

\bibitem[Bernstein(2009)]{Bernstein}
D.~S. Bernstein.
\newblock \emph{Matrix mathematics : theory, facts, and formulas}.
\newblock Princeton University Press, Princeton, N.J. ; Oxford, 2nd ed.
  edition, 2009.
\newblock ISBN 9780691132877.

\bibitem[Bormann et~al.(2011)Bormann, Geer, and Bauer]{bormann11}
N.~Bormann, A.~J. Geer, and P.~Bauer.
\newblock Estimates of observation-error characteristics in clear and cloudy
  regions for microwave imager radiances from numerical weather prediction.
\newblock \emph{Q. J. R. Meteorol. Soc.}, 137:\penalty0 2014--2023, 2011.

\bibitem[Bormann et~al.({2016})Bormann, Bonavita, Dragani, Eresmaa, Matricardi,
  and McNally]{bormann16}
N.~Bormann, M.~Bonavita, R.~Dragani, R.~Eresmaa, M.~Matricardi, and A.~McNally.
\newblock {Enhancing the impact of IASI observations through an updated
  observation error covariance matrix}.
\newblock \emph{Q. J. R. Meteorol. Soc.}, 142\penalty0 (697):\penalty0
  {1767--1780}, {2016}.

\bibitem[Campbell et~al.(2017)Campbell, Satterfield, Ruston, and
  Baker]{campbell16}
W.~F. Campbell, E.~A. Satterfield, B.~Ruston, and N.~L. Baker.
\newblock Accounting for correlated observation error in a dual-formulation
  {4D} variational data assimilation system.
\newblock \emph{Monthly Weather Review}, 145\penalty0 (3):\penalty0 1019--1032,
  2017.
\newblock \doi{10.1175/MWR-D-16-0240.1}.
\newblock URL \url{https://doi.org/10.1175/MWR-D-16-0240.1}.

\bibitem[Chalon et~al.(2001)Chalon, Cayla, and Diebel]{chalon01}
G.~Chalon, F.~Cayla, and D.~Diebel.
\newblock {IASI}: {An} advanced sounder for operational meteorology.
\newblock In \emph{Proceedings of IAF, Toulouse, France}, 2001.

\bibitem[Collard(2007)]{Collard07}
A.~D. Collard.
\newblock Selection of iasi channels for use in numerical weather prediction.
\newblock \emph{Quarterly Journal of the Royal Meteorological Society},
  133\penalty0 (629):\penalty0 1977--1991, 2007.
\newblock \doi{10.1002/qj.178}.
\newblock URL
  \url{https://rmets.onlinelibrary.wiley.com/doi/abs/10.1002/qj.178}.

\bibitem[Collard et~al.(2010)Collard, McNally, Hilton, Healy, and
  Atkinson]{Collard10}
A.~D. Collard, A.~P. McNally, F.~I. Hilton, S.~B. Healy, and N.~C. Atkinson.
\newblock The use of principal component analysis for the assimilation of
  high-resolution infrared sounder observations for numerical weather
  prediction.
\newblock \emph{Quarterly Journal of the Royal Meteorological Society},
  136\penalty0 (653):\penalty0 2038--2050, 2010.
\newblock \doi{10.1002/qj.701}.

\bibitem[Cordoba et~al.(2016)Cordoba, Dance, Kelly, Nichols, and
  Waller]{cordoba16}
M.~Cordoba, S.~L. Dance, G.~A. Kelly, N.~K. Nichols, and J.~A. Waller.
\newblock {Diagnosing Atmospheric Motion Vector observation errors for an
  operational high resolution data assimilation system}.
\newblock \emph{Q. J. R. Meteorol. Soc.}, 2016.
\newblock \doi{10.1002/qj.2925}.

\bibitem[Desroziers et~al.(2005)Desroziers, Berre, Chapnik, and
  Poli]{desroziers05}
G.~Desroziers, L.~Berre, B.~Chapnik, and P.~Poli.
\newblock Diagnosis of observation, background and analysis-error statistics in
  observation space.
\newblock \emph{Q. J. R. Meteorol. Soc.}, 131:\penalty0 3385--3396, 2005.

\bibitem[Eyre(1989)]{Eyre89}
J.~R. Eyre.
\newblock {Inversion of cloudy satellite sounding radiances by nonlinear
  optimal estimation. I: Theory and simulation for TOVS}.
\newblock \emph{Quarterly Journal of the Royal Meteorological Society},
  115\penalty0 (489):\penalty0 1001--1026, 1989.
\newblock \doi{10.1002/qj.49711548902}.

\bibitem[Fowler et~al.(2018)Fowler, Dance, and Waller]{fowler17}
A.~M. Fowler, S.~L. Dance, and J.~A. Waller.
\newblock On the interaction of observation and prior error correlations in
  data assimilation.
\newblock \emph{Quarterly Journal of the Royal Meteorological Society},
  144\penalty0 (710):\penalty0 48--62, 2018.
\newblock \doi{10.1002/qj.3183}.

\bibitem[Gauthier et~al.(2018)Gauthier, Du, Heilliette, and Garand]{Gauthier18}
P.~Gauthier, P.~Du, S.~Heilliette, and L.~Garand.
\newblock Convergence issues in the estimation of interchannel correlated
  observation errors in infrared radiance data.
\newblock \emph{Monthly Weather Review}, 146\penalty0 (10):\penalty0
  3227--3239, 2018.
\newblock \doi{10.1175/MWR-D-17-0273.1}.

\bibitem[Golub and Van~Loan(1996)]{golub96}
G.~H. Golub and C.~F. Van~Loan.
\newblock \emph{Matrix Computations}.
\newblock The John Hopkins University Press, third edition, 1996.

\bibitem[Haben(2011)]{haben11c}
S.~A. Haben.
\newblock \emph{Conditioning and preconditioning of the minimisation problem in
  variational data assimilation}.
\newblock PhD thesis, Department of Mathematics and Statistics, University of
  Reading, 2011.

\bibitem[Haben et~al.(2011)Haben, Lawless, and Nichols]{haben11b}
S.~A. Haben, A.~S. Lawless, and N.~K. Nichols.
\newblock Conditioning of incremental variational data assimilation, with
  application to the {Met} {Office} system.
\newblock \emph{Tellus A}, 64(4):\penalty0 782--792, 2011.

\bibitem[Healy and White(2005)]{healy05}
S.~B. Healy and A.~A. White.
\newblock {Use of discrete Fourier transforms in the 1D-Var retrieval problem}.
\newblock \emph{Quarterly Journal of the Royal Meteorological Society},
  131\penalty0 (605):\penalty0 63--72, 2005.

\bibitem[Hilton et~al.(2009)Hilton, Atkinson, English, and Eyre]{hilton09}
F.~Hilton, N.~C. Atkinson, S.~J. English, and J.~R. Eyre.
\newblock {Assimilation of IASI at the Met Office and assessment of its impact
  through observing system experiments}.
\newblock \emph{Q. J. R. Meteorol. Soc.}, 135:\penalty0 495--505, 2009.

\bibitem[Janji{\'c} et~al.(2018)Janji{\'c}, Bormann, Bocquet, Carton, Cohn,
  Dance, Losa, Nichols, Potthast, Waller, and Weston]{janjic17}
T.~Janji{\'c}, N.~Bormann, M.~Bocquet, J.~A. Carton, S.~E. Cohn, S.~L. Dance,
  S.~N. Losa, N.~K. Nichols, R.~Potthast, J.~A. Waller, and P.~Weston.
\newblock {On the representation error}.
\newblock \emph{Q. J. R. Meteorol. Soc.}, 2018.
\newblock \doi{10.1002/qj.3130}.

\bibitem[Lupu et~al.(2015)Lupu, Cardinali, and McNally]{lupu15}
C.~Lupu, C.~Cardinali, and A.~P. McNally.
\newblock Adjoint-based forecast sensitivity applied to observation-error
  variance tuning.
\newblock \emph{Q. J. R. Meteorol. Soc.}, 141:\penalty0 3157--3165, 2015.

\bibitem[M{\'e}nard(2016)]{menard16}
R.~M{\'e}nard.
\newblock Error covariance estimation methods based on analysis residuals:
  theoretical foundation and convergence properties derived from simplified
  observation networks.
\newblock \emph{Q. J. R. Meteorol. Soc.}, 142:\penalty0 257--273, 2016.

\bibitem[Michel(2018)]{Michel18}
Y.~Michel.
\newblock Revisiting fisher's approach to the handling of horizontal spatial
  correlations of the observation errors in a variational framework.
\newblock \emph{Quarterly Journal of the Royal Meteorological Society}, 2018.
\newblock \doi{10.1002/qj.3249}.
\newblock URL
  \url{https://rmets.onlinelibrary.wiley.com/doi/abs/10.1002/qj.3249}.

\bibitem[Nocedal(2006)]{Nocedal06}
J.~Nocedal.
\newblock \emph{Numerical optimization}.
\newblock Springer series in operations research and financial engineering.
  Springer, New York ; London, 2nd ed. edition, 2006.
\newblock ISBN 9780387303031.

\bibitem[Pavelin and Candy(2014)]{Pavelin14}
E.~G. Pavelin and B.~Candy.
\newblock Assimilation of surface-sensitive infrared radiances over land:
  Estimation of land surface temperature and emissivity.
\newblock \emph{Quarterly Journal of the Royal Meteorological Society},
  140\penalty0 (681):\penalty0 1198--1208, 2014.
\newblock \doi{10.1002/qj.2218}.

\bibitem[Pavelin et~al.(2008)Pavelin, English, and Eyre]{Pavelin08}
E.~G. Pavelin, S.~J. English, and J.~R. Eyre.
\newblock The assimilation of cloud-affected infrared satellite radiances for
  numerical weather prediction.
\newblock \emph{Quarterly Journal of the Royal Meteorological Society},
  134\penalty0 (632):\penalty0 737--749, 2008.
\newblock \doi{10.1002/qj.243}.

\bibitem[Prates et~al.(2016)Prates, Migliorini, Stewart, and Eyre]{prates16}
C.~Prates, S.~Migliorini, L.~Stewart, and J.~Eyre.
\newblock Assimilation of transformed retrievals obtained from clear-sky {IASI}
  measurements.
\newblock \emph{Q. J. R. Meteorol. Soc.}, 142:\penalty0 1697--1712, 2016.

\bibitem[Rainwater et~al.(2015)Rainwater, Bishop, and Campbell]{rainwater15}
S.~Rainwater, C.~H. Bishop, and W.~F. Campbell.
\newblock The benefits of correlated observation errors for small scales.
\newblock \emph{Q.J.R. Meteorol. Soc}, 141:\penalty0 3439--3445, 2015.

\bibitem[Simonin et~al.(2019)Simonin, Waller, Ballard, Dance, and
  Nichols]{simonin18}
D.~Simonin, J.~A. Waller, S.~P. Ballard, S.~L. Dance, and N.~K. Nichols.
\newblock {A pragmatic strategy for implementing spatially correlated
  observation errors in an operational system: an application to Doppler radar
  winds}.
\newblock \emph{Q. J. R. Meteorol. Soc.}, 2019.
\newblock \doi{https://doi.org/10.1002/qj.3592}.

\bibitem[Stewart(2010)]{stewart09}
L.~M. Stewart.
\newblock \emph{Correlated observation errors in data assimilation}.
\newblock PhD thesis, University of Reading, 2010.

\bibitem[Stewart et~al.(2008)Stewart, Dance, and Nichols]{stewart08}
L.~M. Stewart, S.~L. Dance, and N.~K. Nichols.
\newblock Correlated observation errors in data assimilation.
\newblock \emph{Int. J. Numer. Meth.}, 56:\penalty0 1521--1527, 2008.

\bibitem[Stewart et~al.(2009)Stewart, Cameron, Dance, English, Eyre, and
  Nichols]{stewart08b}
L.~M. Stewart, J.~Cameron, S.~L. Dance, S.~English, J.~Eyre, and N.~K. Nichols.
\newblock Observation error correlations in {IASI} radiance data.
\newblock Mathematics report series, 1/2009, University of Reading, Reading,
  UK, 2009.
\newblock URL
  \url{https://www.reading.ac.uk/web/files/maths/obs_error_IASI_radiance.pdf}.

\bibitem[Stewart et~al.(2013)Stewart, Dance, and Nichols]{stewart13}
L.~M. Stewart, S.~L. Dance, and N.~K. Nichols.
\newblock Data assimilation with correlated observation errors: experiments
  with a {1-D} shallow water model.
\newblock \emph{Tellus A}, 65:\penalty0 19546 (14pp), 2013.
\newblock URL \url{http://dx.doi.org/10.3402/tellusa.v65i0.19546}.

\bibitem[Stewart et~al.(2014)Stewart, Dance, Nichols, Eyre, and
  Cameron]{stewart14}
L.~M. Stewart, S.~L. Dance, N.~K. Nichols, J.~R. Eyre, and J.~Cameron.
\newblock {Estimating interchannel observation-error correlations of IASI
  radiance data in the Met Office system}.
\newblock \emph{Q. J. R. Meteorol. Soc.}, 140:\penalty0 1236--1244, 2014.

\bibitem[Tabeart et~al.(2018)Tabeart, Dance, Haben, Lawless, Nichols, and
  Waller]{tabeart17a}
J.~M. Tabeart, S.~L. Dance, S.~A. Haben, A.~S. Lawless, N.~K. Nichols, and
  J.~A. Waller.
\newblock The conditioning of least squares problems in variational data
  assimilation.
\newblock \emph{Numerical Linear Algebra with Applications}, doi:\penalty0
  10.1002/nla.2165 (22pp), 2018.
\newblock URL \url{http://dx.doi.org/10.1002/nla.2165}.

\bibitem[Tabeart et~al.(2019)Tabeart, Dance, Lawless, Nichols, and
  Waller]{tabeart17c}
J.~M. Tabeart, S.~L. Dance, A.~S. Lawless, N.~K. Nichols, and J.~A. Waller.
\newblock {Improving the conditioning of estimated covariance matrices}.
\newblock 2019.
\newblock Submitted.

\bibitem[Waller et~al.(2014)Waller, Dance, Lawless, Nichols, and
  Eyre]{waller14}
J.~A. Waller, S.~L. Dance, A.~S. Lawless, N.~K. Nichols, and J.~R. Eyre.
\newblock Representativity error for temperature and humidity using the {Met}
  {Office} high-resolution model.
\newblock \emph{Q.J.R. Meteorol. Soc}, 140:\penalty0 1189--1197, 2014.

\bibitem[Waller et~al.(2016{\natexlab{a}})Waller, Ballard, Dance, Kelly,
  Nichols, and Simonin]{waller16b}
J.~A. Waller, S.~P. Ballard, S.~L. Dance, G.~Kelly, N.~K. Nichols, and
  D.~Simonin.
\newblock {Diagnosing Horizontal and Inter-Channel Observation Error
  Correlations for SEVIRI Observations Using Observation-Minus-Background and
  Observation-Minus-Analysis Statistics}.
\newblock \emph{Remote Sensing}, 8 (7):\penalty0 851, 2016{\natexlab{a}}.
\newblock URL \url{10.3390/rs8070581}.

\bibitem[Waller et~al.(2016{\natexlab{b}})Waller, Dance, and Nichols]{waller16}
J.~A. Waller, S.~L. Dance, and N.~K. Nichols.
\newblock Theoretical insight into diagnosing observation error correlations
  using observation-minus-background and observation-minus-analysis statistics.
\newblock \emph{Q.J.R. Meteorol. Soc}, 142:\penalty0 418--431,
  2016{\natexlab{b}}.

\bibitem[Waller et~al.(2016{\natexlab{c}})Waller, Simonin, Dance, Nichols, and
  Ballard]{waller16c}
J.~A. Waller, D.~Simonin, S.~L. Dance, N.~K. Nichols, and S.~P. Ballard.
\newblock {Diagnosing Observation Error Correlations for Doppler Radar Radial
  Winds in the Met Office UKV Model Using Observation-Minus-Background and
  Observation-Minus-Analysis Statistics}.
\newblock \emph{{Monthly Weather Review}}, {144}\penalty0 ({10}):\penalty0
  {3533--3551}, 2016{\natexlab{c}}.

\bibitem[Waller et~al.(2017)Waller, Dance, and Nichols]{waller17}
J.~A. Waller, S.~L. Dance, and N.~K. Nichols.
\newblock On diagnosing observation-error statistics with local ensemble data
  assimilation.
\newblock \emph{Quarterly Journal of the Royal Meteorological Society},
  143\penalty0 (708):\penalty0 2677--2686, 2017.
\newblock \doi{10.1002/qj.3117}.

\bibitem[Wang et~al.(2018)Wang, Fei, Cheng, Huang, and Zhong]{Wang2018}
T.~Wang, J.~Fei, X.~Cheng, X.~Huang, and J.~Zhong.
\newblock Estimating the correlated observation-error characteristics of the
  chinese fengyun microwave temperature sounder and microwave humidity sounder.
\newblock \emph{Advances in Atmospheric Sciences}, 35\penalty0 (11):\penalty0
  1428--1441, 2018.
\newblock ISSN 1861-9533.
\newblock \doi{10.1007/s00376-018-8014-9}.

\bibitem[Weston(2011)]{weston11}
P.~Weston.
\newblock Progress towards the implementation of correlated observation errors
  in {4D-Var}.
\newblock Forecasting research technical report 560, Met Office, Exeter, UK,
  2011.

\bibitem[Weston et~al.(2014)Weston, Bell, and Eyre]{weston14}
P.~P. Weston, W.~Bell, and J.~R. Eyre.
\newblock Accounting for correlated error in the assimilation of
  high-resolution sounder data.
\newblock \emph{Q.J.R. Meteorol. Soc}, 140:\penalty0 2420--2429, 2014.

\end{thebibliography}
\end{document}